\documentclass[%
preprint,
nofootinbib,
amsmath,
amssymb,
aps,
prd,
]{revtex4-2}

\usepackage{graphicx}
\usepackage{bm}
\usepackage{amsfonts}

\usepackage{xcolor}

\DeclareFontFamily{U}{mathx}{}
\DeclareFontShape{U}{mathx}{m}{n}{<-> mathx10}{}
\DeclareSymbolFont{mathx}{U}{mathx}{m}{n}
\DeclareMathAccent{\widehat}{0}{mathx}{"70}
\DeclareMathAccent{\widecheck}{0}{mathx}{"71}

\renewcommand{\tilde}{\widetilde}
\renewcommand{\hat}{\widehat}
\newcommand{\re}[1] {(\ref{#1})}
\newcommand{\chk}{\widecheck}
\renewcommand{\bar}{\overline}
\newcommand{\tr}{{\rm Tr}}
\newcommand{\dg}{\dagger}
\newcommand{\cpone}{\mathbb{CP}^1}
\newcommand{\triR}{\hbox{\,\Large$\triangleright$}}
\newcommand{\triL}{\hbox{\,\Large$\triangleleft$}}

\newcommand{\nn}{\nonumber}

\begin{document}


\title{New techniques for Gauge Theories in Projective Superspace}

\author{Ariunzul Davgadorj}
\email{ariunzul.d@gmail.com}
 \affiliation{Department of Theoretical Physics and Astrophysics, Faculty of Science, Masaryk University, Kotlářská 2, 61137 Brno, Czechia.}
\author{Ulf Lindstr\"om}%
 \email{ulf.lindstrom@physics.uu.se}
\affiliation{%
 Theoretical Physics, Imperial College. Prince Consort Rd, London SW7 2AZ, UK.\\
 and\\
 Department of Physics and Astronomy, Theoretical Physics, Uppsala University SE 751 20 Uppsala, Sweden.
}%

\author{Rikard von Unge}
 \email{unge@physics.muni.cz}
\affiliation{Department of Theoretical Physics and Astrophysics, Faculty of Science, Masaryk University, Kotlářská 2, 61137 Brno, Czechia.
}%

\date{\today}

\begin{abstract}
We introduce new techniques for calculations in Gauge theories with extended supersymmetry. We are working in Projective Superspace where the $SU(2)$ R-symmetry is realized geometrically by including an auxilliary $\cpone$ component in the superspace. Different gauge representations are associated with different dependence on the $\cpone$ coordinate $\zeta$ and using contour integrals on $\cpone$ we define natural projection operators on these different representations which leads to elegant formulas for all relevant objects. The new techniques lead to compact expressions for lagrangians and field strengths in terms of the gauge prepotential but also to effective ways of reducing superspace expressions to components, i.e. to write them in terms of fields transforming covariantly only under a subgroup of the supersymmetry group. We illustrate our findings in several examples in three and four dimensions.
\end{abstract}

\maketitle

\tableofcontents

\section{Introduction}
Projective superspace\footnote{Not to be confused with the generalization of ordinary projective spaces to the superworld} \cite{Karlhede:1984vr} is a formalism where theories with a supersymmetry algebra with an R-symmetry group containing one or several factors of $SU(2)$ can be treated keeping the supersymmetry manifest. Because of its close connection to the twistor space description of Hyperk\"ahler manifolds \cite{Hitchin:1986ea,Karlhede:1986mg,Lindstrom:2009afn} and Quaternion K\"ahler manifolds \cite{deWit:2001brd,Anguelova:2004sj} it has over the years led to many interesting applications in both mathematics and physics. Among others, one could mention the Hyperk\"ahler quotient construction \cite{Lindstrom:1983rt,Hitchin:1986ea}, the explanation of the wall crossing phenomena \cite{Gaiotto:2010okc} and the c-map \cite{Gates:1999zv,Rocek:2005ij,Rocek:2006xb}. In physics, projective superspace makes it possible to calculate contributions to the effective action keeping the supersymmetry manifest. This technique was established for Hypermultiplet contributions in \cite{Gonzalez-Rey:1997pxs,Gonzalez-Rey:1997msl}. Even more interesting are the vector multiplet contributions where, for a long time, only the Abelian case was fully known. The topic of Yang-Mills theory in projective superspace was introduced in \cite{Lindstrom:1989ne} and further developments can be found in \cite{Gonzalez-Rey:1997spr}. In a previous publication \cite{Davgadorj:2017ezp} we made some contributions to this topic by finding explicit expressions for the field strengths in terms of the gauge prepotential $e^V$ and showing that they have the expected properties. Also, supergravity with extended supersymmetry has been described in the projective superspace formalism \cite{Kuzenko:2007cj,Kuzenko:2007hu,Kuzenko:2008wr,Kuzenko:2008ep,Kuzenko:2009zu,Butter:2014xxa}. For a review we recommend \cite{Kuzenko:2010bd}.

The types of theories that can be described in Projective superspace can also be described in Harmonic superspace, albeit not always very directly, See \cite{Kuzenko:1998xm,Butter:2012ta,Butter:2015nza}. In particular, Yang-Mills theory in Harmonic superspace was treated in \cite{Galperin:1984av}. A nice introduction to the topic can be found in \cite{Galperin:2001seg}.

In both Harmonic and Projective superspace, the $SU(R)_R$ symmetry is kept manifest by introducing an auxilliary $\cpone = \frac{SU(2)}{U(1)}$ on which all the superfields depend so that $SU(2)_R$ transformations are realized as coordinate transformations on this $\cpone$. In the Harmonic approach the superfields depend on the full $SU(2)$ harmonics and the $U(1)$ dependence is factored out by considering fields with fixed $U(1)$ charge and actions that are $U(1)$ invariant. In contrast, the Projective superspace formalism has factored out the $U(1)$ dependence from the start by only considering superfields that depend holomorphically on $\cpone$. Contour integrals on $\cpone$ are then used to construct invariant actions. A careful comparison between the Harmonic and Projective superspace formalisms can be found in \cite{Kuzenko:1998xm,Butter:2012ta}.

A hybrid formalism between Projective and Harmonic superspace, called Hyperspace, has also been developed in \cite{Jain:2009aj,Jain:2010gm,Jain:2012jx,Jain:2012zx,Jain:2013hua}. In \cite{Davgadorj:2017ezp} we took the results of \cite{Jain:2012zx} to develop the Projective superspace formalism for Yang-Mills theory. A key insight used in this as well as in our previous paper is that all objects should be expanded in powers of $X = e^V-1$, where $V$ is the gauge prepotential, taken at different points of $\cpone$. Furthermore, at several points in our calculations it is convenient to write our objects in terms of objects regular around the north pole or south pole of the auxiliary $\cpone$. This may be done using projection operators defined by contour integrals modified with the $\epsilon$-prescription introduced in \cite{Jain:2009aj}. It is interesting that the projection operators defined in this way do not project symmetrically on positive and negative powers of $\zeta$ but nevertheless lead to elegant formulas that can be used in further calculations.

The paper is organized as follows: in section \ref{sec:PSS} we introduce the Projective superspace formalism, in particular we discuss general aspects of gauge theory in the Projective superspace approach. In section \ref{sec:NewTech} we introduce the new approach by expanding all objects of interest in terms of $X=e^V-1$ and using projectors defined using the $\epsilon$-prescription. We try to keep the discussion as general as possible. After this general discussion, in order to illustrate the power of the new techniques introduced in this paper, we treat several specific examples in section \ref{sec:Ex}. We treat gauge theories with eight supercharges in three and four dimensions and also with six supercharges in three dimensions. In the latter case we find the surprising fact that the Lagrangian that gives Yang-Mills theory for the eight supercharge theories gives Chern-Simons theory in the case with six supercharges. In appendix \ref{app:eps} we review the $\epsilon$-prescription.

\section{Gauge Theory in Projective superspace\label{sec:PSS}}
\subsection{Projective superspace}
Projective superspace is available as soon as there is an $SU(2)$ R-symmetry in the theory. The R-symmetry is kept manifest by introducing an auxilliary projective space, $\cpone$, with coordinate $\zeta$. All superfields transform as sections of some holomorphic line bundle over this $\cpone$.

The basic matter multiplet is a superfield $\Upsilon(\zeta)$, analytic around the north pole of the $\cpone$ so it can be expanded as
\begin{align}
\Upsilon(\zeta) = \sum_{n=0}^{\infty} \Upsilon_n \zeta^n ~.
\end{align}
Fields with this $\zeta$ dependence are called Arctic multiplets. There are also fields that are analytic around the south pole of the $\cpone$, they can be expanded as
\begin{align}
\widetilde{\Upsilon}(\zeta) = \sum_{n=-\infty}^{0}\widetilde{\Upsilon}_n \zeta^{n}~,
\end{align}
and are called Antarctic multiplets.

Further, there are  multiplets with polynomial dependence on $\zeta$. The ${\mathcal O}(2k)$ multiplet is defined as
\begin{align}
\eta(\zeta) = \sum_{n=-k}^k \eta_n \zeta^n~.
\end{align}
Of particular importance is the case where $k\rightarrow\infty$ which is called the Tropical multiplet since it is singular at both the north- and the south-pole.

Conjugation of a generic superfield $T(\zeta) = \sum_{n} t_n\zeta^n$ is given by hermitian conjugation of the coefficients $t_n$ combined with the antipodal map on the $\cpone$
\begin{align}\label{eq:conjugation}
\bar{T}(\zeta) = \bar{\sum_n t_n\zeta^n} = \sum_n t^\dg_n \left(-\frac{1}{\zeta}\right)^n~.
\end{align}

Using the conjugation it makes sense to define self-conjugate (or "real") ${\mathcal O}(2k)$ multiplets by imposing $\bar{\eta} = \eta$ which implies
\begin{align}
\sum_{n=-k}^k \eta_n\zeta^n = \sum_{n=-k}^k (-1)^n \eta_{-n}^\dg \zeta^n~.
\end{align}
This makes sense also for the Tropical multiplet where $k\rightarrow \infty$.

It is also possible to define $\zeta$ dependent supercovariant derivatives that transform as sections of a line bundle of $\cpone$. The particular line bundle (and thus the $\zeta$ dependence) is given by the representation of the derivatives under the R-symmetry. In the four dimensional theory with eight supercharges, there are two $SU(2)_R$ doublets of supercovariant derivatives $D^i_\alpha$ and $\bar{D}_{i\dot\alpha}$. They transform in the fundamental and antifundamental representation of $SU(2)_R$. Therefore, the definition of the {\em projective} supercovariant derivatives becomes
\begin{align}\nn
\nabla_\alpha &= D^1_\alpha + \zeta D^2_\alpha\\
\label{Na2Def}
\bar{\nabla}_{\dot\alpha} &= \bar{D}_{2\dot\alpha} -\zeta \bar{D}_{1\dot\alpha} ~,
\end{align}
and they anticommute among themselves. In the general discussion of this paper, we will often illustrate the new techniques on examples involving this particular situation, i.e. on the case of $\mathcal{N}=2$ supersymmetric gauge theory in four dimensions. However, our results are more generally true. In other cases the projective derivatives look slightly different (see for instance equation (\ref{eq:N3nabla}) or (\ref{eq:N4nabla})) but the main conclusions remain valid.

Since the supercovariant derivatives constructed in this way anticommute with each other they can be used to define constrained superfields.
In the ${\mathcal N} =2$ case defined above the supercovariant projective derivatives (\ref{Na2Def}) can be used to define constrained superfields $\Upsilon$ as the kernel of this set of operators
\begin{align}
\nabla_\alpha \Upsilon = 0~ , \;\; \bar{\nabla}_{\dot\alpha}\Upsilon = 0~.
\end{align}
We call such fields {\em projective} superfields\footnote{Sometimes also referred to as projectively chiral superfields.}. The most common instance is if the field $\Upsilon$ is complex and has arctic $\zeta$ dependence. It is then a projective version of the hypermultiplet, or some generalization thereof. In this case we also have the conjugate field $\bar{\Upsilon}$ which is projective but with antarctic $\zeta$ dependence. The action is given by
\begin{align}\label{eq:Ssimp}
\int d^4 x d^4 \theta \oint \frac{d\zeta}{2\pi i \zeta} \bar{\Upsilon}\Upsilon
\end{align}
where the integral is over only half of the superspace since the fields are constrained. In what follows, we will suppress the $2\pi i$ factor in the measure.

\subsection{Gauge transformations and Gauge invariance}

In this section we mimic the usual superspace gauging of super Yang-Mills as far as possible.

Given an arctic projecive field it can be transformed by a global phase rotation where $\Lambda$ is a constant
\begin{align}
 \Upsilon  \rightarrow e^{i\Lambda} \Upsilon\label{YP}~.
\end{align}
and the action (\ref{eq:Ssimp}) is invariant under this global symmetry.
To make this transformation local, we allow the parameter $\Lambda$ to become a superfield itself. In order to respect the properties of $\Upsilon$ we need $\Lambda$ to also be projective with arctic $\zeta$ dependence. Similarly we have
\begin{align}
\bar{\Upsilon} \rightarrow \bar{\Upsilon} e^{-i\bar{\Lambda}}\label{YPB}~,
\end{align}
where $\bar{\Lambda}$ is the conjugate of $\Lambda$ and thus projective with antarctic $\zeta$ dependence.
The superspace lagrangian
\begin{align}
\bar{\Upsilon}\Upsilon~,
\end{align}
which is invariant under the global phase rotation, is not invariant in the local case. In order to make it invariant under the local transformation we introduce the projective superspace gauge field $V(\zeta)$ which is projective, self conjugate and with tropical $\zeta$ dependence. Under gauge transformations it transforms as
\begin{align}
e^V \rightarrow e^{i\bar{\Lambda}} e^V e^{-i\Lambda}~,
\end{align}
so that the expression
\begin{align}
\bar{\Upsilon} e^V \Upsilon
\end{align}
is invariant.
Since $V$ and the gauge parameters are projective superfields, the supercovariant derivatives $\nabla,\bar{\nabla}$ are automatically also gauge covariant without adding any connection coefficients. Using $e^V$ we may redefine fields to transform with only the arctic parameter $\Lambda$, or, alternatively with the antarctic parameter $\bar{\Lambda}$ only. This we call the arctic/antarctic representation where derivatives of gauge covariant fields are also gauge covariant.

Alternatively one may split $e^V$ into parts with only positive or negative powers of $\zeta$
\begin{align}
 e^V = e^{\bar{U}}e^U~,
\end{align}
where $e^{U}$ contains only non-negative powers of $\zeta$ and its conjugate $e^{\bar{U}}$ contains only non-positive powers. Under gauge transformations they transform
\begin{align}\nn
e^U &\rightarrow e^{iK} e^U e^{-i\Lambda}\\
e^{\bar{U}} &\rightarrow e^{i\bar{\Lambda}} e^{\bar{U}} e^{-iK}~,
\end{align}
where $K$ is a real $\zeta$ independent superfield. Using $e^U$ and $e^{\bar{U}}$ we may convert all fields in the theory to transform with the parameter $K$ only. This we call the vector representation. The transformation of the covariant derivatives in going from the polar to the vector representation is given by a similarity transformation
\begin{align}\label{eq:sim}
[\nabla_\alpha]_{v} = e^U \nabla_\alpha e^{-U} = e^{-\bar{U}} \nabla_\alpha e^{\bar{U}} =\nabla_\alpha+\Gamma_\alpha~.
\end{align}
The second equality follows from the fact that $V$ is projective since then
\begin{align}
0 = \nabla_\alpha e^V = \nabla_\alpha (e^{\bar{U}}e^{U}) = 
(\nabla_\alpha e^{\bar{U}})e^{U} + e^{\bar{U}} (\nabla_\alpha e^{U})~.
\end{align}
We thus find that in the arctic/antarctic representations the supercovariant derivatives do not develop any nontrivial connection coefficients and they stay the same as the supercovariant derivatives in the ungauged theory. This should be contrasted to the situation in the vector representation in which the gauged supercovariant derivaties do have a nontrivial connection coefficient piece.

\subsection{The field strength}\label{FS1}
When we look at the anticommutation relations between the projective covariant derivatives in the vector representation, taken at different points on the $\cpone$, $\zeta_1$ and $\zeta_2$  say, they no longer anticommute. In all cases studied in this paper the anticommutator is proportional to a field strength
\begin{align}
\{[\nabla_{\alpha}]_v (\zeta_1) , [\nabla_{\beta}]_v(\zeta_2) \} = - (\zeta_2-\zeta_1)  C_{\alpha\beta} {\mathbb W}^\dg~,
\end{align}
which, introducing a derivative $\frac{\partial}{\partial \zeta}$ on the $\cpone$, allows us to write
\begin{align}\label{anticom}
\left\{ [\nabla_{\alpha}]_v ,
\left[\frac{\partial}{\partial\zeta} ,
[\nabla_{\beta}]_v \right] \right\} &= - C_{\alpha\beta} {\mathbb W}^\dg~,
\end{align}
using the definition of the $\zeta$ deivative.
Multiplying the relation \re{anticom} by appropriate exponential factors while using the similarity transformation (\ref{eq:sim}) taking us back to the arctic representation where we have
\begin{align}
\left\{\nabla_{\alpha},
\left[e^{-U}\frac{\partial}{\partial\zeta} e^{U},
\nabla_{\beta} \right] \right\} &= - C_{\alpha\beta} e^{-U} {\mathbb W}^\dg e^{U} = - C_{\alpha\beta} {\mathcal W}^\dg~.
\end{align}
Here we have introduced the notation $\mathbb{W},\mathbb{W}^\dg$ for the field strength in the vector representation and $\mathcal{W},\mathcal{W}^\dg$ for the field strength in the arctic representation (and $\tilde{\mathcal{W}},\tilde{\mathcal{W}}^\dg$ in the antarctic case). Later we will also need $W,W^\dg,W^0$ as $\zeta$ components of field strengths that explicitly depend on $\zeta$. Notice that we use $\dg$ to denote the conjugate field strength since the bar denotes the projective superspace conjugation (\ref{eq:conjugation}) that also changes the representation from arctic to antarctic. 

Defining the gauge covariant $\zeta$-derivative with gauge field $\mathcal A$, that gauges the symmetry \re{YP},\re{YPB},
\begin{align}\label{eq:Aarctic}
{\mathcal D}_\zeta = \partial_\zeta + {\mathcal A} = e^{-U}\partial_\zeta e^{U}~,\end{align}
we see that the field strength in the Arctic representation may be calculated as
\begin{align}\label{eq:fs}
{\mathcal W}^\dg = - \nabla^2 {\mathcal A}~.
\end{align}
Similar formulas can be derived in the antarctic representation where the gauge field is given by
\begin{align}\label{eq:Aantarctic}
 \tilde{\mathcal A} = e^{\bar{U}} (\partial_\zeta e^{-\bar{U}})~,
\end{align}

\subsection{Solving for ${\mathcal A}$}
In \cite{Davgadorj:2017ezp} we expressed $\mathcal A$ in terms of the gauge potential $V$ by writing it as an expansion in powers of $X = e^V-1$. Using (\ref{eq:Aarctic}) and (\ref{eq:Aantarctic}) we wrote
\begin{align}
\partial_\zeta X = {\mathcal A} - \tilde{\mathcal A} + X{\mathcal A} -\tilde{\mathcal A}X~,
\end{align}
`which gave a recursion relation in powers of $X$ that can be solved. To find the arctic representation potential ${\mathcal A}$ or the antarctic representation potential $\tilde{\mathcal A}$ one needs to project onto positive or negative powers of $\zeta$. This can be elegantly done using contour integrals and the $\epsilon$-prescription\footnote{This is a procedure for avoiding poles analogous to the usual $i\epsilon$ procedure in QFT.} introduced in \cite{Jain:2009aj,Jain:2010gm} (see Appendix \ref{app:eps} for more details). In short, we use the notation
\begin{align}
\frac{1}{\zeta_{10}} &\equiv 
\frac{1}{\zeta_1} \sum_{n=0}^{\infty} \left(\frac{\zeta_0}{\zeta_1}\right)^n 
\left(= \frac{1}{\zeta_1-\zeta_0} \;\;\;{\rm if}\;\;\; \left|\frac{\zeta_0}
{\zeta_1}\right|<1\right) \\ \nonumber
\frac{1}{\zeta_{01}} &\equiv
\frac{1}{\zeta_0} \sum_{n=0}^{\infty} \left(\frac{\zeta_1}{\zeta_0}\right)^n \left(=
\frac{1}{\zeta_0-\zeta_1} \;\;\;{\rm if} \;\;\; \left|\frac{\zeta_1}
{\zeta_0}\right|<1 \right)~.
\end{align}
Using this, in \cite{Davgadorj:2017ezp} we found that, after some non obvious manipulations, the gauge potential (\ref{eq:Aarctic}) can be written
\begin{align}\label{eq:zconn}
{\mathcal A}(\zeta_0) = \sum_{n=1}^{\infty} (-1)^{n+1} \oint d\zeta_1 \ldots d\zeta_n \frac{1}{\zeta_{10}}
\frac{X_1\ldots X_n}{\zeta_{21}\ldots\zeta_{n,n-1}} \frac{1}{\zeta_{n0}}~,
\end{align}
where $X_k = e^{V(\zeta_k)} - 1$. We also found that under gauge transformations, $\mathcal A$ transforms as a gauge field
\begin{align}
\delta{\mathcal A} = - i \partial_\zeta \Lambda + [i\Lambda , {\mathcal A}]
= -i[{\mathcal D}_\zeta,\Lambda]~.
\end{align}
It is also interesting to observe that under conjugation, the arctic representation gauge field ${\mathcal A}$ transforms into the corresponding field in the antarctic representation (\ref{eq:Aantarctic})
\begin{align}
\bar{\mathcal A} = -\zeta^2\tilde{\mathcal A}~.
\end{align}

\section{New Techniques\label{sec:NewTech}}
The analysis up until now has been made by splitting the gauge field into parts with only non-negative or non-positive powers of $\zeta$.
\begin{align}\label{eq:split}
e^V = e^{\bar{U}}e^{U}~,
\end{align}
in such a way that $e^U$ and $e^{\bar{U}}$ are conjugates of each other and subsequently expanding everything in powers of $X = e^V-1$. This turns out to be the wrong choice if we would like to find useful expressions for $e^U$. The main problem is that since $e^U$ and $e^{\bar{U}}$ are conjugates of each other and, in particular, contain equal parts of terms independent of $\zeta$, it is awkward to write the resulting expressions using the $\epsilon$-prescription which naturally puts the $\zeta$ independent terms together with the terms with positive powers of $\zeta$ (\ref{eq:eppr}).

\subsection{The splitting}
To illustrate this, let us formally expand $e^U$ in powers of $X = e^V-1$
\begin{align}\label{Y}
e^{U} &= 1 + Y^{(1)} + Y^{(2)} + \ldots \\
\label{YY}
e^{\bar{U}} &= 1 + \bar{Y}^{(1)} + \bar{Y}^{(2)} + \ldots
\end{align}
where $Y^{(n)}$ contains $n$ powers of $X$. Inserting this ansatz into (\ref{eq:split}) we get
\begin{align}
e^V = 1+X = \Bigl(1 + \bar{Y}^{(1)} + \bar{Y}^{(2)} + \ldots \Bigr)
\Bigl(1 + Y^{(1)} + Y^{(2)} + \ldots \Bigr)~,
\end{align}
which gives us an infinite set of equations that can be solved recursively,
\begin{align}
\bar{Y}^{(1)} + Y^{(1)} &= X \nonumber \\ 
\bar{Y}^{(2)} + Y^{(2)} &= -\bar{Y}^{(1)} Y^{(1)} \\
\bar{Y}^{(3)} + Y^{(3)} &= -\bar{Y}^{(2)} Y^{(1)} - \bar{Y}^{(1)} Y^{(2)}\nonumber \\
& \;\; \vdots\nonumber 
\end{align}
At each stage of the solution we have to project the expression on the right hand side on positive or negative powers of $\zeta$ with a symmetric splitting of the $\zeta$ independent terms so that $\bar{Y}^{(n)}$ is the conjugate of $Y^{(n)}$ in agreement with \re{Y} and \re{YY}.

If we instead use projection operators defined using contour integrals and the $\epsilon$-prescription as in (\ref{eq:eppr}) and (\ref{eq:epdef}), it naturally favors an asymmetrical projection where the $\zeta$ independent part is put together with the terms with positive powers of $\zeta$
\begin{align}\label{eq:projec}
\oint d\zeta_1 \frac{X_1}{\zeta_{10}} &= \sum_{n=0}^{\infty} x_n \zeta_0^n \\
\oint d\zeta_1 \frac{X_1}{\zeta_{01}} &= \sum_{n=-\infty}^{-1} x_n \zeta_0^n \nonumber
\end{align}
We therefore make a different ansatz, writing $e^V = e^{\chk{\bar{U}}} e^{\hat{U}}$ where now $e^{\hat{U}}$ contains all the $\zeta$ independent terms. Expanding in powers of $X$ we get 
\begin{align}
e^{\hat{U}} &= 1+ \hat{Y}^{(1)} + \hat{Y}^{(2)} + \dots \\
e^{\chk{\bar{U}}} &= 1+ \chk{\bar{Y}}^{(1)} + \chk{\bar{Y}}^{(2)} + \dots
\end{align}
which leads to the recursive relations
\begin{align}\label{eq:Yrec}
\chk{\bar{Y}}^{(1)} + \hat{Y}^{(1)} &= X  \nonumber \\
\chk{\bar{Y}}^{(2)} + \hat{Y}^{(2)} &= -\chk{\bar{Y}}^{(1)} \hat{Y}^{(1)} \\ \nonumber
\chk{\bar{Y}}^{(3)} + \hat{Y}^{(3)} &= -\chk{\bar{Y}}^{(2)} \hat{Y}^{(1)} - \chk{\bar{Y}}^{(1)} \hat{Y}^{(2)} \\ \nonumber
& \;\; \vdots 
\end{align}
Solving (\ref{eq:Yrec}) and implementing the projections through the contour integrals and the $\epsilon$ prescription we find that $e^{\hat{U}}$ and $e^{\chk{\bar{U}}}$ can be written elegantly as
\begin{align}\label{eq:Uhat}
e^{\hat{U}(\zeta_0)} &= 1+ \sum_{n=1}^\infty (-1)^{n+1} \oint d\zeta_1 \ldots d\zeta_n \frac{X_1\ldots X_n}{\zeta_{21} \ldots \zeta_{n,n-1}}\frac{1}{\zeta_{n0}} \\
\label{eq:bUchk}
e^{\chk{\bar{U}}(\zeta_0)} &= 1+ \sum_{n=1}^{\infty} (-1)^{n+1} \oint d\zeta_1\ldots d\zeta_n
 \frac{1}{\zeta_{01}}\frac{X_1\ldots X_n}{\zeta_{21} \ldots \zeta_{n,n-1}}~.
\end{align}

It is also possible to choose a projection where the $\zeta$ independent terms are put together with the negative powers of $\zeta$ since
\begin{align}\label{eq:opprojec}
\oint d\zeta_1 \frac{X_1}{\zeta_{10}} \frac{\zeta_0}{\zeta_1} &= \sum_{n=1}^{\infty} x_n \zeta_0^n \\
\oint d\zeta_1 \frac{X_1}{\zeta_{01}} \frac{\zeta_0}{\zeta_1} &= \sum_{n=-\infty}^{0} x_n \zeta_0^n~.
\nonumber
\end{align}
In this case we are instead lead to the ansatz $e^V = e^{\hat{\bar{U}}}e^{\chk{U}}$ and the expressions
\begin{align}\label{eq:Uchk}
e^{\chk{U}(\zeta_0)} &= 1+ \sum_{n=1}^\infty (-1)^{n+1} \oint d\zeta_1 \ldots d\zeta_n \frac{\zeta_0}{\zeta_1} \frac{X_1\ldots X_n}{\zeta_{21} \ldots \zeta_{n,n-1}} \frac{1}{\zeta_{n0}}\\
\label{eq:bUhat}
e^{\hat{\bar{U}}(\zeta_0)} &= 1+ \sum_{n=1}^{\infty} (-1)^{n+1} \oint d\zeta_1\ldots d\zeta_n
 \frac{1}{\zeta_{01}}\frac{X_1\ldots X_n}{\zeta_{21} \ldots \zeta_{n,n-1}} \frac{\zeta_0}{\zeta_n}
\end{align}

For completeness we also give the inverse expressions
\begin{align}
\label{eq:mUhat}
e^{-\hat{U}(\zeta_0)} &= 1+ \sum_{n=1}^\infty (-1)^{n} \oint d\zeta_1 \ldots d\zeta_n \frac{1}{\zeta_{10}}\frac{X_1\ldots X_n}{\zeta_{21} \ldots \zeta_{n,n-1}} \\
\label{eq:mUchk}
e^{-\chk{U}(\zeta_0)} &= 1+ \sum_{n=1}^\infty (-1)^{n} \oint d\zeta_1 \ldots d\zeta_n \frac{1}{\zeta_{10}}\frac{X_1\ldots X_n}{\zeta_{21} \ldots \zeta_{n,n-1}} \frac{\zeta_0}{\zeta_n}\\
\label{eq:mbUhat}
e^{-\hat{\bar{U}}(\zeta_0)} &= 1+ \sum_{n=1}^{\infty} (-1)^n \oint d\zeta_1\ldots d\zeta_n
\frac{\zeta_0}{\zeta_1} \frac{X_1\ldots X_n}{\zeta_{21} \ldots \zeta_{n,n-1}} \frac{1}{\zeta_{0n}}\\
\label{eq:mbUchk}
e^{-\chk{\bar{U}}(\zeta_0)} &= 1+ \sum_{n=1}^{\infty} (-1)^n \oint d\zeta_1\ldots d\zeta_n \frac{X_1\ldots X_n}{\zeta_{21}\ldots\zeta_{n,n-1}}
\frac{1}{\zeta_{0n}}
\end{align}

\bigskip
We are now in the situation that we may write
\begin{align}\label{splits}
e^V = e^{\bar{U}}e^U = e^{\chk{\bar{U}}}e^{\hat{U}} = e^{\hat{\bar{U}}} e^{\chk{U}}~,
\end{align}
where the expressions differ in where we have put the $\zeta$ independent terms according to the chosen projection, (\ref{eq:projec}) or (\ref{eq:opprojec}). We may formally isolate the $\zeta$ independent part that is different in $e^{\hat{U}}$, $e^{U}$ and $e^{\chk{U}}$ and write $e^{\hat{U}} = e^P e^{U}$. That $e^P$ is $\zeta$ independent can be seen from the fact that using (\ref{splits}) we may write $e^P = e^{\hat{U}} e^{-U} = e^{-\chk{\bar{U}}}e^{\bar{U}}$ explicitly showing that $e^P$ contains at the same time only non-negative and only non-positive powers of $\zeta$. From the definition also follows that $e^U = e^{\bar{P}} e^{\chk{U}}$ so that $e^{\hat{U}} = e^P e^{U} = e^P e^{\bar{P}} e^{\chk{U}}$.

We may express $e^P e^{\bar{P}} = e^{\hat{U}}e^{-\chk{U}}= e^{-\chk{\bar{U}}}e^{\hat{\bar{U}}}$ using contour integrals as
\begin{align}\label{eq:PPbar}
e^P e^{\bar{P}} = 1+\sum_{n=1}^\infty (-1)^{n+1} \oint d\zeta_1 \ldots d\zeta_n \frac{X_1\ldots X_n}{\zeta_{21}\ldots\zeta_{n,n-1}} \frac{1}{\zeta_n}~,
\end{align}
with inverse
\begin{align}\label{eq:PbarP}
e^{-\bar{P}} e^{-P} = 1+\sum_{n=1}^\infty (-1)^{n} \oint d\zeta_1 \ldots d\zeta_n \frac{1}{\zeta_1} \frac{X_1\ldots X_n}{\zeta_{21}\ldots\zeta_{n,n-1}}~,
\end{align}
which can be written compactly defining the Hermitian and $\zeta$ independent quantities
\begin{align}\label{eq:Ps}
\Gamma^{(n)} &= \oint d\zeta_1 \ldots d\zeta_n \frac{X_1\ldots X_n}{\zeta_{21}\ldots\zeta_{n,n-1}} \frac{1}{\zeta_n}~,\nonumber \\
L^{(n)} &= \oint d\zeta_1 \ldots d\zeta_n \frac{1}{\zeta_1} \frac{X_1\ldots X_n}{\zeta_{21}\ldots\zeta_{n,n-1}}~,
\end{align}
as
\begin{align}
e^P e^{\bar{P}} &= 1+\sum_{n=1}^\infty (-1)^{n+1} \Gamma^{(n)}~,\nonumber \\
e^{-\bar{P}} e^{-P} &= 1+\sum_{n=1}^\infty (-1)^{n} L^{(n)}~.
\end{align}
Using the rules for conjugation of the fields including the contour integrals and the $\epsilon$-regulated $\zeta$-denominators given in Appendix \ref{app:eps} we can check that $\Gamma^{(n)}$ and $L^{(n)}$ are self conjugate and furthermore that
\begin{align}
L^{(n)} - \Gamma^{(n)} = \sum_{k=1}^{n-1} \Gamma^{(k)} L^{(n-k)} = \sum_{k=1}^{n-1} L^{(k)}\Gamma^{(n-k)}~.
\end{align}
This we show by writing the left hand side using (\ref{eq:Ps}) as
\begin{align}
L^{(n)} - \Gamma^{(n)} = \oint d\zeta_1 \ldots d\zeta_n \frac{X_1\ldots X_n}{\zeta_{21}\ldots\zeta_{n,n-1}} \left(\frac{1}{\zeta_1} -  \frac{1}{\zeta_n} \right)~,
\end{align}
and then rewriting the $\zeta$ factors in two different ways
\begin{align}
\frac{1}{\zeta_1}-\frac{1}{\zeta_n} &= \frac{1}{\zeta_1\zeta_n} \left[(\zeta_n-\zeta_{n-1})+\ldots+(\zeta_2-\zeta_1)\right]~,\\ \nonumber
\frac{1}{\zeta_1}-\frac{1}{\zeta_n} &= \frac{\zeta_n-\zeta_{n-1}}{\zeta_{n-1}\zeta_n} +\ldots+
\frac{\zeta_2 - \zeta_1}{\zeta_1\zeta_2}~,
\end{align}
and then canceling the $\zeta$-factors $(\zeta_k-\zeta_{k-1})$ that appears with the $\epsilon$-regulated $\frac{1}{\zeta_{k,k-1}}$ factors in the numerator. That this is possible is again proven in Appendix \ref{app:eps}. We conclude that (\ref{eq:PPbar}) and (\ref{eq:PbarP}) are in fact inverses of each other.

In the following, it will also be useful to introduce the additional $\zeta$ independent but Hermitian conjugate pair
\begin{align}\label{eq:Adef}
A^{(n)} &= \oint d\zeta_1\ldots d\zeta_n \frac{X_1\ldots X_n}{\zeta_{21}\ldots \zeta_{n,n-1}} ~,\nonumber \\
\bar{A}^{(n)} &= -\oint d\zeta_1\ldots d\zeta_n \frac{1}{\zeta_1}\frac{X_1\ldots X_n}{\zeta_{21}\ldots \zeta_{n,n-1}} \frac{1}{\zeta_n}~.
\end{align}

\bigskip
Notice that since $e^P$ is $\zeta$ independent we may use any representation to calculate the connection coefficient of the $\zeta$ derivative
\begin{align}
{\cal A} = e^{-U}\partial_\zeta e^{U} = e^{-\hat{U}} \partial_\zeta e^{\hat{U}} = e^{-\chk{U}} \partial_\zeta e^{\chk{U}}~,
\end{align}
which can also be checked explicitly 
using the expressions (\ref{eq:Uhat}),(\ref{eq:Uchk}),(\ref{eq:mUhat}) and (\ref{eq:mUchk}).

It is also interesting to check how (\ref{eq:PPbar}) and (\ref{eq:PbarP}) transform under gauge transformations. An explicit calculation gives
\begin{align}
\delta\left(e^Pe^{\bar{P}}\right) &= i\bar{\lambda}_0 e^P e^{\bar{P}} - e^Pe^{\bar{P}}i\lambda_0~,\\ \nonumber
\delta\left(e^{-\bar{P}}e^{-P}\right) &= i{\lambda}_0 e^{-\bar{P}}e^{-P} - e^{-\bar{P}}e^{-P}i\bar{\lambda}_0~,
\end{align}
where $\lambda_0$ and $\bar{\lambda}_0$ are the $\zeta$-independent pieces of $\Lambda(\zeta)$ and $\bar{\Lambda}(\zeta)$ and are precisely the gauge parameters of the $\zeta$-{\em components} of the superfield. More precisely, the gauge field $V$ transforms under gauge transformations as
\begin{align}
\delta V = i\bar{\Lambda}-i\Lambda +\mathcal{O}(V)~.
\end{align}
This implies the $\zeta$-component transformations
\begin{align}
\delta v_0 &= i\bar{\lambda}_0 - i\lambda_0 + \mathcal{O}(v)~.
\end{align}
This allows us to identify $v_0$ as the component gauge field with the standard component gauge transformation with $\lambda_0,\bar{\lambda}_0$ as the parameter.

Given the definition of $e^P = e^{\hat{U}}e^{-U}$ and the transformation properties of $e^U \rightarrow e^{iK}e^U e^{-i\Lambda}$ we find that $e^P$ and $e^{\bar{P}}$ must transform as
\begin{align}
e^P &\rightarrow e^{i\bar{\lambda}_0} e^P e^{-iK}~,\\ \nonumber
e^{\bar{P}} &\rightarrow e^{iK}e^{\bar{P}}e^{-i\lambda_0}~,
\end{align}
and furthermore that
\begin{align}
e^{\hat{U}} &\rightarrow e^{i\bar{\lambda}_0} e^{\hat{U}}e^{-i\Lambda}~,\\
\nonumber
e^{\chk{U}} &\rightarrow e^{i\lambda_0} e^{\chk{U}} e^{-i\Lambda}~,
\end{align}
which again can be confirmed by explicit calculation. This gives a nice interpretation of the fields $e^{\hat{U}}, e^{\chk{U}}, e^P, e^{\bar{P}}$ as converters between the arctic, vector, chiral/component and antichiral/component representations.

\subsection{The basic potentials}
To illustrate the power of our new formalism we will calculate $\zeta$-independent expressions for the connection coefficients of the four dimensional (vector representation) gauge covariant derivatives as defined in (\ref{eq:sim})
\begin{align}\nonumber
\Gamma_\alpha &= e^U(\nabla_\alpha e^{-U}) = e^{-P}e^{\hat{U}} \nabla_\alpha(e^{-\hat{U}}e^P) =
e^{-P}(\nabla_\alpha e^P) + e^{-P}e^{\hat{U}} (\nabla_\alpha e^{-\hat{U}})e^P
\\
& = e^{\bar{P}}e^{\chk{U}}\nabla_\alpha ( e^{-\chk{U}}e^{-\bar{P}}) =
e^{\bar{P}}\nabla_\alpha e^{-\bar{P}} + e^{\bar{P}}e^{\chk{U}} (\nabla_\alpha e^{-\chk{U}})e^{-\bar{P}}~,
\end{align}
where $\nabla_\alpha  = D^1_\alpha + \zeta D^2_{\alpha}
=:D_\alpha + \zeta Q_\alpha$, thus defining $D_\alpha$ and $Q_\alpha$. (See section \ref{sec:Ex} for a thorough discussion of the four dimensional case.)
We would like to calculate $\nabla_\alpha e^{-\chk{U}}$. For notational simplicity, let us denote the $\zeta$ coordinate on which they depend by $\zeta_0$. Then we need to calculate
\begin{align}\nonumber
\nabla_{0\alpha} e^{-\chk{U}(\zeta_0)} &=
\nabla_{0\alpha} \sum_{n=1}^\infty (-1)^{n} \oint d\zeta_1 \ldots d\zeta_n \frac{1}{\zeta_{10}}\frac{X_1\ldots X_n}{\zeta_{21} \ldots \zeta_{n,n-1}} \frac{\zeta_0}{\zeta_n}
\\
\label{eq:concoff}
&=
\sum_{n=1}^\infty (-1)^{n} \oint d\zeta_1 \ldots d\zeta_n 
\sum_{k=1}^{n}\frac{1}{\zeta_{10}}\frac{X_1\ldots \nabla_{0\alpha}X_k \ldots X_n}{\zeta_{21} \ldots \zeta_{n,n-1}} \frac{\zeta_0}{\zeta_n}~.
\end{align}
We may use that $X = e^V -1$ is a projective superfield to write
\begin{align}\label{eq:projX}
\nabla_{0\alpha} X_k = \left(1-\frac{\zeta_0}{\zeta_k}\right) D_\alpha X_k =
\left(\zeta_0 - \zeta_k\right) Q_\alpha X_k~,
\end{align}
By expanding the $\zeta$ factors at hand as
\begin{align}\label{eq:expleftQ}
(\zeta_0 - \zeta_k) &= (\zeta_0 - \zeta_1) + (\zeta_1 - \zeta_2)
+ \ldots + (\zeta_{k-1}-\zeta_k)
\end{align}
if we decide to expand using $Q_\alpha$, or
\begin{align}
\left(1-\frac{\zeta_0}{\zeta_k}\right) &=
\frac{\zeta_0(\zeta_1-\zeta_0)}{\zeta_0\zeta_1}
+
\frac{\zeta_0(\zeta_2-\zeta_1)}{\zeta_1\zeta_2}
+ \ldots +
\frac{\zeta_0(\zeta_k-\zeta_{k-1})}{\zeta_{k-1}\zeta_k}\label{eq:expleftD}~,
\end{align}
if we decide to expand using $D_\alpha$. Then
canceling the $\epsilon$-regulated $\zeta$-factors in the denominator of (\ref{eq:concoff}) with the $\zeta$-factors from (\ref{eq:expleftQ}) or (\ref{eq:expleftD}) we find a number of relations which we may reorganize to read
\begin{align}
\Gamma_\alpha = e^{\bar{P}}D_\alpha e^{-\bar{P}}+\zeta e^{-P}Q_\alpha e^P~,
\end{align}
as well as
\begin{align}
e^{-P}Q_\alpha e^P &= e^{\bar{P}} Q_\alpha e^{-\bar{P}}
- e^{\bar{P}} D_\alpha \bar{A} e^{-\bar{P}}~,\\ \nonumber
e^{\bar{P}} D_\alpha e^{-\bar{P}} &= e^{-P} D_\alpha e^{P}
+ e^{-P} Q_\alpha A e^P~,
\end{align}
where we have defined
\begin{align}\label{eq:Apotdef}
A &= \sum_{n=1}^{\infty} (-1)^{n+1} A^{(n)}\nonumber \\
\bar{A} &= \sum_{n=1}^{\infty} (-1)^{n+1}\bar{A}^{(n)}~,
\end{align}
using the definition (\ref{eq:Adef}).

Again it is useful to look at how these fields transform under gauge transformations. An explicit calculation gives
\begin{align}
\delta A &= i\bar{\lambda}_0 A - A i\bar{\lambda}_0 - i\bar{\lambda}_{-1}~,\\
\nonumber
\delta \bar{A} &= i\lambda_0 \bar{A} -\bar{A} i\lambda_0 - i \lambda_1~.
\end{align}
From the projectiveness of $\Lambda$ and $\bar{\Lambda}$ we find that
\begin{align}
D_\alpha \lambda_0 = \bar{D}_\alpha \bar{\lambda}_0 = 0~,\\ \nonumber
D^2 \lambda_1 = \bar{D}^2 \bar{\lambda}_1 = 0~,
\end{align}
which means that $D^2\bar{A}$ and $\bar{D}^2A$ transform covaraintly as
\begin{align}
D^2 \bar{A} &\rightarrow e^{i\lambda_0} D^2 \bar{A} e^{-i\lambda_0}~,\\
\nonumber
\bar{D}^2 A &\rightarrow e^{i\bar{\lambda}_0} \bar{D}^2 A e^{-i\bar{\lambda}_0}~,
\end{align}
appropriate for the component field strength in the chiral or antichiral representation. This will be confirmed in the next section.

\subsection{The field strength\label{sec:FSintro}}
Similarly we may calculate expressions for the field strength (\ref{eq:fs}) from the $\zeta$-connection coefficient (\ref{eq:zconn})
\begin{align}
{\mathcal W}_0^\dg = -\nabla_0^2 {\mathcal A}_0 = -\sum_{n=1}^{\infty} (-1)^{n+1}
\oint d\zeta_1\ldots d\zeta_n
\frac{1}{\zeta_{10}}
\frac{\nabla_{0}^2 [X_1\ldots X_n]}{\zeta_{21}\ldots\zeta_{n,n-1}}
\frac{1}{\zeta_{n0}}~.
\end{align}
We let the projective derivatives act and like in (\ref{eq:projX}) use that $X$ is projective.
When we let the derivatives act there will be terms of the type
\begin{align}
X_1\ldots \nabla_0^\alpha X_l \ldots \nabla_{0\alpha} X_r \ldots X_n~,
\end{align}
as well as terms where both derivatives act on the same $X$
\begin{align}\label{eq:QuadD}
X_1\ldots \nabla_0^2X_k\ldots X_n~.
\end{align}
For the leftmost derivative we rewrite the $\zeta$-factor as in (\ref{eq:expleftQ}) or (\ref{eq:expleftD}) but the for the derivative on the right we instead use
\begin{align}\label{eq:exprightQ}
(\zeta_0 - \zeta_r) &= (\zeta_0-\zeta_n) + (\zeta_n - \zeta_{n-1})
+ \ldots +
(\zeta_{r+1}-\zeta_r)~,
\end{align}
in the case we expand using $Q_\alpha$, or
\begin{align}\label{eq:exprightD}
\left(1-\frac{\zeta_0}{\zeta_r}\right) &= \frac{\zeta_0(\zeta_n-\zeta_0)}{\zeta_0\zeta_n}+ \frac{\zeta_0(\zeta_{n-1}-\zeta_n)}{\zeta_{n-1}\zeta_n}
+ \ldots +
\frac{\zeta_0(\zeta_r-\zeta_{r+1})}{\zeta_r\zeta_{r+1}}~,
\end{align}
in the case we expand using $D_\alpha$.
For the terms of type (\ref{eq:QuadD}) we expand one of the $\zeta$ factors according to (\ref{eq:expleftQ}) or (\ref{eq:expleftD}) but the other one as in (\ref{eq:exprightQ}) or (\ref{eq:exprightD}).
If we again cancel the differences of $\zeta$-factors in the numerator with the $\epsilon$-regulated $\zeta$-factors in the denominator we find that each term in the field strength can be written as a product of three factors which sum up to give
\begin{align}
{\cal W}^\dg = e^{-\chk{U}} (D^2 \bar{A}) e^{\chk{U}} = e^{-\hat{U}}(Q^2 A) e^{\hat{U}}~.
\end{align}
All of the $\zeta$ dependence is contained in the outer factors and could be removed by a similarity transformation. In a trace with similar objects, these factors drop out. Furthemore we can find the expression for the field strength in the vector representation by
\begin{align}
{\mathbb W}^\dg &= e^U {\mathcal W}^\dg e^{-U} 
= e^{U}e^{-\chk{U}} (D^2 \bar{A}) e^{\chk{U}} e^{-U}  = 
e^{\bar{P}}(D^2 \bar{A}) e^{-\bar{P}} \nonumber \\
&= e^{U} e^{-\hat{U}}(Q^2 A) e^{\hat{U}} e^{-U} =
e^{-P} (Q^2 A) e^{P}~,
\end{align}
which is manifestly $\zeta$ independent.

\subsection{The Lagrangian}\label{Lag1}
The superspace Lagrangian is given by the interesting expression
\begin{align}\label{eq:genFI}
{\cal L} = \sum_{n=1}^{\infty} \frac{(-1)^{n}}{n}
   \tr \oint d\zeta_1\ldots d\zeta_n
    \frac{X_1\ldots X_n}{\zeta_{21}\ldots\zeta_{n,n-1}\zeta_{1n}}~,
\end{align}
which was shown to be gauge invariant in \cite{Jain:2012zx}. It can be thought of as a generalized Fayet-Iliopoulos term since if we would ignore the zeta dependence the sum could be performed with the result $\ln(1+X) = \ln(e^V) = V$. However, the nontrivial $\zeta$ dependence makes the behavior much more interesting as we shall see in the next section.

It is also interesting to investigate how this term transforms under gauge transformations. Since the basic field $X$ transforms as
\begin{align}
\delta X = i\bar{\Lambda} - i\Lambda + i\bar{\Lambda}X - X i\Lambda
\end{align}
one finds that there has to be cancellation between terms of different order of $X$. In particular one finds that at order $X^0$, the term transforms as an ordinary FI-term, with a "K\"ahler gauge transformation".

\section{Examples\label{sec:Ex}}

\subsection{${\cal N}=2$ in four dimensions}
\subsubsection{The algebra}
In $4d$ the ${\cal N} = 2$ superspace has an $SU(2)$ R-symmetry. The covariant derivatives transform in the fundamental and antifundamental representation of the R-symmetry. This leads to the algebra
\begin{align}
\{{\mathcal D}^i_\alpha, {\mathcal D}^k_\beta\} &= -\epsilon^{ik} C_{\alpha\beta} {\mathbb W}^\dg ~,\nonumber \\
\{\bar{\mathcal D}_{i\dot\alpha}, \bar{\mathcal D}_{k\dot\beta}\} &= \epsilon_{ik} C_{\dot\alpha\dot\beta} {\mathbb W}~,\\ \nonumber
\{{\mathcal D}^i_\alpha, \bar{\mathcal D}_{k\dot\beta}\} &= i\delta^i_k \nabla_{\alpha\dot\beta}~.
\end{align}
From the Bianchi identities we learn that
\begin{align}
\bar{\mathcal D}_{i\dot\alpha}{\mathbb W} = {\mathcal D}^i_\alpha {\mathbb W}^\dg = 0~.
\end{align}
Introducing the projective derivatives
\begin{align}
\nabla_{\alpha} = D^1_\alpha +\zeta D^2_\alpha =:D_\alpha +\zeta Q_\alpha ~,\;\;\;
~~~~~\bar{\nabla}_{\dot\alpha} = \bar{D}_{2\dot\alpha} - \zeta \bar{D}_{1\dot\alpha}=:\bar{Q}_{\dot\alpha} - \zeta \bar{D}_{\dot\alpha}~,
\end{align}
analytic around the north pole of $\cpone$ and anticommuting among themselves, we may introduce constrained superfields by
\begin{align}
\nabla_{\alpha} \Upsilon = \bar{\nabla}_{\dot\alpha} \Upsilon = 0~.
\end{align}
As mentioned below \re{Na2Def},
such fields will be called projective superfields. Although the projective derivatives anticommute taken at the same point in $\zeta$, when we take the anticommutator between projective derivatives at different $\zeta$ coordinates, they do not anticommute any more (c.f. \re{FS1})
\begin{align}
\{\nabla_{\alpha}(\zeta_1) , \nabla_{\beta}(\zeta_2) \} &= -C_{\alpha\beta}(\zeta_2-\zeta_1){\mathbb W}^\dg ~,\\ \nonumber
\{\bar{\nabla}_{\dot\alpha}(\zeta_1) , \bar{\nabla}_{\dot\beta}(\zeta_2) \} &=
C_{\dot{\alpha}\dot{\beta}}(\zeta_2-\zeta_1){\mathbb W}~,
\end{align}
which gives us the possibility to express the field strength by introducing the $\zeta$ derivative as
\begin{align}
\{\nabla_{\alpha} , [\frac{\partial}{\partial\zeta} , \nabla_{\beta} ]\} &= - C_{\alpha\beta} {\mathbb W}^\dg~,\\ \nonumber
\{\bar{\nabla}_{\dot\alpha} , [\frac{\partial}{\partial\zeta} , \bar{\nabla}_{\dot\beta} ]\} 
&=  C_{\dot{\alpha}\dot{\beta}} {\mathbb W} ~.
\end{align}

\subsubsection{The field strength}
As we have seen, this leads to the field strength in the Arctic representation given by
\begin{align}
{\cal W} &= \bar{\nabla}^2 {\cal A} ~,\\ \nonumber
{\cal W}^\dg &= -\nabla^2 {\cal A}~,
\end{align}
and, as we showed in section \ref{sec:FSintro}, we can write
\begin{align}
{\cal W} &= e^{-\hat{U}} (\bar{D}^2 A )e^{\hat{U}}  = e^{-\chk{U}} (\bar{Q}^2 \bar{A}) e^{\chk{U}} = e^{-U} e^{-P} (\bar{D}^2 A ) e^P e^U~,\\ \nonumber
{\cal W}^\dg &= e^{-\chk{U}} (D^2 \bar{A}) e^{\chk{U}} = e^{-\hat{U}}(Q^2 A) e^{\hat{U}} = e^{-U} e^{\bar{P}} (D^2 \bar{A}) e^{-\bar{P}}e^U~,
\end{align}
where $A$ and its conjugate $\bar{A}$ are $\zeta$ independent fields defined in (\ref{eq:Apotdef}).
As shown in section \ref{sec:FSintro}, the field strength in the vector representation can be written as
\begin{align}\label{Wbold}
\mathbb{W} &= e^{-P} (\bar{D}^2 A ) e^P ~,\\ \nonumber
{\mathbb{W}}^\dg & = e^{\bar{P}} (D^2 \bar{A}) e^{-\bar{P}}~.
\end{align}

\subsubsection{The action}
The action is the generalized Fayet-Iliopoulos term we introduced in section \ref{Lag1}. It is integrated over the full superspace which in this case is
\begin{align}
\int d^4 x d^8\theta = \int d^4x D^2Q^2 \bar{D}^2 \bar{Q}^2~.
\end{align}
We may write the final $\bar{Q}^2$ in the measure as $D^2Q^2 \bar{D}^2 \bar{Q}^2 = D^2Q^2 \bar{D}^2 (\bar{Q} - \zeta \bar{D})^2 = D^2Q^2 \bar{D}^2 \bar{\nabla}^2$ since the terms that were added are dependent on $\bar{D}$ and are projected out by the rest of the measure. We may also choose the $\zeta$ coordinate of the $\nabla^2$ operator in the measure arbitrarily since the terms containing $\zeta$ are projected out in the same way.
Taking this into account and choosing the $\zeta$ in $\nabla$ to be $\zeta_1$ we write the action as
\begin{align}
S & = 
\sum_{n=2}^{\infty}\frac{(-1)^n}n \int d^4 x \oint d\zeta_1 \ldots d\zeta_n D^2 Q^2 \bar{D}^2 \bar{\nabla}_1^2
\; \tr \left [ \frac{X_1 X_2\ldots X_n}{\zeta_{21}\ldots\zeta_{1n}} \right] \\ \nonumber
&=
\sum_{n=2}^{\infty}\frac{(-1)^n}n \int d^4 x \oint d\zeta_1 \ldots d\zeta_n D^2 Q^2  \bar{D}^2
\; \tr \left [\frac{X_1 \bar{\nabla}_1^2 (X_2\ldots X_n)}{\zeta_{21}\ldots\zeta_{1n}} \right] \; .
\end{align}
Concentrating on the expression in the trace we act with the $\bar{\nabla}_1$ derivatives and, analogously to the procedure in section \ref{sec:FSintro}, use that on any projective field $\bar{\nabla}_1 X_k = (\zeta_k-\zeta_1)\bar{D}X_k$
\begin{align}
&\tr \left [\frac{X_1 \bar{\nabla}_1^2 (X_2\ldots X_n)}{\zeta_{21}\ldots\zeta_{1n}} \right] \;  
\nonumber \\  & = 
\sum_{l=2}^{n-1} \sum_{r=l+1}^n (\zeta_l -\zeta_1)(\zeta_r-\zeta_1) \;\tr \left [\frac{X_1\ldots \bar{D}^{\dot\alpha}X_l\ldots \bar{D}_{\dot\alpha} X_r \ldots X_n}{\zeta_{21}\ldots\zeta_{1n}} \right] \;   \label{eq:n2action} \\ & +
\sum_{k=2}^{n} (\zeta_k -\zeta_1)^2 \;\tr \left [\frac{X_1\ldots \bar{D}^2 X_k \ldots \ldots X_n}{\zeta_{21}\ldots\zeta_{1n}} \right] \;. \nonumber
\end{align}
Following the procedure described in section \ref{sec:FSintro} we then rewrite $(\zeta_l-\zeta_1) = (\zeta_l-\zeta_{l-1}) + \ldots + (\zeta_2-\zeta_1)$ and $(\zeta_r-\zeta_1) = (\zeta_r-\zeta_{r+1}) + \ldots + (\zeta_n-\zeta_1)$ as well as $(\zeta_k-\zeta_1)^2 = \left[(\zeta_k-\zeta_{k-1}) + \ldots + (\zeta_2-\zeta_1)\right] \times \left[(\zeta_k-\zeta_{k+1}) + \ldots + (\zeta_n-\zeta_1)\right]$.
Canceling the $\zeta$ factors in the numerator with the factors in the denominator and using the cyclicity of the trace we get
\begin{align}\label{rawF}
S &= \sum_{n=2}^{\infty}(-1)^{n+1} \int d^4 x D^2 Q^2 \bar{D}^2
\; \tr \sum_{k=1}^{n-1} \frac{k}{n} A^{(k)}\bar{D}^2 A^{(n-k)}~.
\end{align}
Integrating by parts, the expression (\ref{rawF}) can be rewritten
\begin{align}
S &= \sum_{n=2}^{\infty}(-1)^{n+1} \int d^4 x D^2 Q^2 \bar{D}^2
\; \tr \sum_{k=1}^{n-1} \frac{n-k}{n} A^{(k)}\bar{D}^2 A^{(n-k)}~.
\end{align}
Averaging over the two alternative expressions we get
\begin{align}
S &= \frac12 \sum_{n=2}^{\infty}(-1)^{n+1} \int d^4 x D^2 Q^2\bar{D}^2
\; \tr \sum_{k=1}^{n-1} A^{(k)}\bar{D}^2 A^{(n-k)} \nonumber \\ &=
-\frac12  \int d^4 x D^2 Q^2\bar{D}^2
\; \tr A \bar{D}^2 A
=
-\frac12  \int d^4 x D^2 Q^2
\; \tr \bar{D}^2 A \bar{D}^2 A~,\label{eq:n2YM}
\end{align}
where $A = \sum_{k=1}^\infty (-1)^{k+1} A^{(k)}$ as defined before. This is indeed the Yang-Mills action
\begin{align}\label{eq:chiralYM}
-\frac12  \int d^4 x D^2 Q^2
\; \tr \mathbb{W}\mathbb{W} = -\frac12 \int d^4 x D^2 Q^2 \oint \frac{d\zeta}{\zeta} \tr \mathcal{W}\mathcal{W}~,
\end{align}
with ${\mathbb W} = \frac12 e^{-P} (\bar{D}^2 A) e^{P}$ and ${\mathcal W} = e^{-\hat{U}} (\bar{D}^2 A )e^{\hat{U}}$. In \cite{Davgadorj:2017ezp} we presented a proof of the equivalence of the Yang-Mills action in the form (\ref{eq:genFI}) and (\ref{eq:chiralYM}) which was rather contrived. Using the new methods introduced in this paper we have presented a straightforward proof.

\subsection{${\cal N}=3$ in three dimensions}
This theory has been studied before in Harmonic superspace by Zupnik \cite{Zupnik:1997nn} and in Projective superspace in \cite{Arai:2011fi,Arai:2012dn,Arai:2013eta} albeit in an ${\mathcal N} = 3$ unitary gauge. 

\subsubsection{The algebra}
The $\mathcal {N}=3$ superspace has $SO(3)$ R-symmetry where the three supercovariant derivatives transform as a triplet. We index them with a symmetric pair of fundamental $SU(2)$ indices. Including a Yang-Mills field and using the symmetry properties of the covariant derivatives, we can write down the most general algebra which gives the algebra of the gauge super covariant derivtives
\begin{align}
    \{{\mathcal D}^{(ik)}_\alpha , {\mathcal D}^{(lm)}_\beta\} = \left(\epsilon^{il}\epsilon^{km}+\epsilon^{im}\epsilon^{kl}\right)i\nabla_{\alpha\beta} +
    \frac{i}{2}C_{\alpha\beta}\left(\epsilon^{il}W^{km}+\epsilon^{kl}W^{im}+\epsilon^{im}W^{kl}+\epsilon^{km}W^{il}\right)
\end{align}
where the field strengths $W^{ik}$ is again given by an $SU(2)_R$ triplet of fields encoded by a symmetric pair of fundamental $SU(2)$ indices. Renaming the derivatives as $\mathcal{D}^{11}\equiv\mathcal{D},\;\;\mathcal{D}^{22}\equiv\bar{\mathcal{D}}\;\;$and\;\;$i\mathcal{D}^{12}\equiv\mathcal{D}^{0}$ and the field strengths as $W^{11}\equiv {W}^\dg,\;\; W^{22}\equiv W$ and $iW^{12}\equiv W^0$ we get
\begin{align}
\nonumber
    &\left\{\mathcal{D}_\alpha,\bar{\mathcal{D}}_\beta\right\} =
    2 i\nabla_{\alpha\beta}  + 2 C_{\alpha\beta} W^0~,\\
    &\left\{\mathcal{D}^{0}_\alpha,\mathcal{D}^{0}_\beta\right\} =
     i\nabla_{\alpha\beta} ~,\\
     \nonumber
    &\left\{\mathcal{D}^0_\alpha,\mathcal{D}_\beta\right\} = C_{\alpha\beta}{W}^\dg ~,\\
    \nonumber
    &\left\{\bar{\mathcal{D}}_\alpha,\mathcal{D}^0_\beta\right\} = C_{\alpha\beta}W~.
\end{align}
As a consequence of the Bianchi identities we then have additional conditions for the field strengths
\begin{align}\label{bianchi}
\nonumber
    & \mathcal{D}_\alpha {W}^\dg=0=\bar{\mathcal{D}}_{\alpha} W \qquad (chirality) ~,\\
    & \mathcal{D}_\alpha ^{0} {W}^\dg=-\mathcal{D}_\alpha W^0 ~,\\
    \nonumber
    & \mathcal{D}_\alpha ^{0} W=-\bar{\mathcal{D}}_{\alpha}W^0 ~,\\
    \nonumber
    & \mathcal{D}_\alpha ^{0} W^0=\frac{1}{4}\left(\mathcal{D}_\alpha W+\bar{\mathcal{D}}_{\alpha} {W}^\dg\right)~,
\end{align}
which further imply that
\begin{align}
    {\mathcal D}^2 W^0 = \bar{\mathcal D}^2 W^0 = 0~,
\end{align}
meaning that $W^0$ is a real linear superfield from the point of view of the ${\cal N} =2$ superfield components.

These relations clearly show that the supercovariant derivatives can be put together into a projective supercovariant derivative that transforms as an ${\mathcal O}(2)$ multiplet
\begin{align}\label{eq:N3nabla}
\nabla_\alpha(\zeta) = \frac{1}{\zeta}{\mathcal D}_\alpha + 2{\mathcal D}^0_\alpha -\zeta \bar{\mathcal D}_\alpha~,
\end{align}
and anticommute among themselves
\begin{align}
\{\nabla_\alpha(\zeta),\nabla_\beta(\zeta) \} = 0~.
\end{align}
Similarly the three field strengths can also be put into an ${\mathcal O}(2)$ multiplet
\begin{align}\label{eq:n3fs}
{\mathbb W} = \frac{1}{\zeta} {W}^\dg +2W^0 - \zeta W~.
\end{align}
Using this description, the Bianchi identities can be compactly written
\begin{align}
\nabla_\alpha {\mathbb W} = 0~,
\end{align}
i.e. by saying that the field strength is Projective. This theory has been discussed before in Projective superspace \cite{Arai:2011fi,Arai:2012dn,Arai:2013eta} where they however were working in $\mathcal{N}=3$ unitary gauge where all the components of $V(\zeta)$ except for $v_{-1},v_0$ and $v_1$ are gauged away.

\subsubsection{The field strength}
The projective supercovariant derivatives anticommute even when the algebra is gauged. To get nontrivial results one needs to look at the anti commutators between derivatives at different $\zeta$ points
\begin{align}\nonumber
\{ \nabla_\alpha(\zeta_1),\nabla_\beta(\zeta_2) \} &= -\frac{(\zeta_1-\zeta_2)^2}{\zeta_1\zeta_2}
2i\nabla_{\alpha\beta}\\ &+2C_{\alpha\beta}
\left(\left(\frac{1}{\zeta_2}-\frac{1}{\zeta_1}\right) \bar{W} + \left(\frac{\zeta_1}{\zeta_2}-\frac{\zeta_2}{\zeta_1}\right)W^0 + \left(\zeta_2-\zeta_1\right) W
\right)
\end{align}
which, by introducing the $\zeta$ derivative, can be used to write the field strength as
\begin{align}
C_{\alpha\beta}\frac{2}{\zeta}{\mathbb W}(\zeta) = -\{\nabla_\alpha,[\partial_\zeta,\nabla_\beta]\}~,
\end{align}
or, by going to the arctic representation as
\begin{align}\label{eq:N3FS}
\boldsymbol{\mathcal W}(\zeta) = -\frac{\zeta}{2}\nabla^2 {\mathcal A}
= \frac{1}{\zeta}{\mathcal W}^\dg +2{\mathcal W}^0 - \zeta{\mathcal W}~,
\end{align}
where again ${\mathcal A} = e^{-U}\partial_\zeta e^{U}$.

We may now calculate explicit expressions for the field strength using this formula. Here we will use that when acting on a projective superfield we can write
\begin{align}\label{eq:N3comp}
\nabla_{0\alpha} X_1 = \frac12\left[
(\zeta_1-\zeta_0)\bar{D}_\alpha X_1 + \frac{\zeta_1-\zeta_0}{\zeta_0\zeta_1} D_\alpha X_1
\right]~.
\end{align}
We now calculate the field strength from (\ref{eq:N3FS}) using (\ref{eq:zconn})
\begin{align}
\boldsymbol{\mathcal W}_0 = -\frac{\zeta_0}{2}\nabla_0^2 {\mathcal A}_0 = 
-\frac{\zeta_0}{2}\sum_{n=1}^{\infty} (-1)^{n+1}
\oint d\zeta_1\ldots d\zeta_n \tr\left[
\frac{1}{\zeta_{10}}
 \frac{\nabla_{0}^2 [X_1\ldots X_n]}{\zeta_{21}\ldots\zeta_{n,n-1}}
\frac{1}{\zeta_{n0}}\right]~.
\end{align}
Concentrating on the expression in the trace we get
\begin{align}\nonumber
\nabla_0^2\tr\left[\frac{1}{\zeta_{10}}
\frac{X_1\ldots X_n}{\zeta_{21}\ldots\zeta_{n,n-1}}
\frac{1}{\zeta_{n0}}
\right] &=
\sum_{l=1}^{n-1}\sum_{r=l+1}^n \tr\left[
\frac{1}{\zeta_{10}}
\frac{X_1\ldots\nabla_0^\alpha X_l \ldots \nabla_{0\alpha} X_r\ldots X_n}
{\zeta_{21}\ldots\zeta_{n,n-1}}
\frac{1}{\zeta_{n0}}\right] \\
&+
\sum_{k=1}^{n} \tr\left[
\frac{1}{\zeta_{10}}
\frac{X_1\ldots\nabla_0^2 X_k \ldots X_n}
{\zeta_{21}\ldots\zeta_{n,n-1}}
\frac{1}{\zeta_{n0}}\right]~.
\end{align}
Using (\ref{eq:N3comp}) there will be three different types of contributions with different $\zeta_0$ dependence. One with two $D$ derivatives, one with two $\bar{D}$ derivatives and a mixed one with one $D$ and one $\bar{D}$. The terms with two $\bar{D}$ or two $D$ operators are treated completely analogously to what we did before. Identifying components by comparing the powers of $\zeta$ in (\ref{eq:N3FS}) the result is
\begin{align}\nn
{\mathcal W} &= \frac{1}{2} e^{-\hat{U}}\bar{D}^2 A \;e^{\hat{U}} ~,\\
{\mathcal W}^\dg &= \frac{1}{2} e^{-\chk{U}} D^2\bar{A} \; e^{\chk{U}}~.
\end{align}
This result is given in the Arctic representation. Going back to the vector representation by conjugating with $e^U$ we get
\begin{align}\nn
W &= \frac12 e^{-P}\bar{D}^2 A \; e^{P} ~,\\
{W}^\dg &= \frac{1}{2} e^{\bar{P}} D^2 \bar{A} e^{-\bar{P}}~.
\end{align}
The mixed term is harder to analyze. The $\zeta$ factors will be either
\begin{align}
(\zeta_l - \zeta_0)\times \frac{\zeta_r - \zeta_0}{\zeta_0\zeta_r}~,
\end{align}
if the $\bar{D}$ operator is on the left of the $D$ operator or
\begin{align}
\frac{\zeta_l - \zeta_0}{\zeta_0\zeta_l}\times(\zeta_r - \zeta_0)~,
\end{align}
in the opposite case. In both cases we rewrite
\begin{align}\label{eq:n3exp}
(\zeta_l-\zeta_0)(\zeta_r-\zeta_0) = \left[(\zeta_l-\zeta_{l-1}) + \ldots + (\zeta_1-\zeta_0)\right] \times
\left[(\zeta_r-\zeta_{r+1}) + \ldots + (\zeta_n-\zeta_0)\right]
\end{align}
and we also always expand
\begin{align}
\frac{1}{\zeta_l} = \frac{\zeta_{l+1}-\zeta_l}{\zeta_l\zeta_{l+1}} + \ldots + \frac{\zeta_f-\zeta_{f-1}}{\zeta_{f-1}\zeta_f}+ \frac{1}{\zeta_f}~,\\
\nonumber
\frac{1}{\zeta_r} = \frac{\zeta_{r+1}-\zeta_r}{\zeta_r\zeta_{r+1}} + \ldots + \frac{\zeta_f-\zeta_{f-1}}{\zeta_{f-1}\zeta_f} + \frac{1}{\zeta_f}~,
\end{align}
where $\zeta_f$ is the $\zeta$ coordinate of the $X_f$ where the denominator $\frac{1}{\zeta_{f+1,f}}$ was canceled by a $(\zeta_{f+1}-\zeta_f)$ from the expansion (\ref{eq:n3exp}).
Keeping track of signs and factors, this can be summed to
\begin{align}
{\mathcal W}^0 = \frac12 e^{-\hat{U}} \bar{D}^\alpha \left( D_\alpha (e^{P}e^{\bar{P}}) e^{-\bar{P}}e^{-{P}}\right) e^{\hat{U}} = \frac12 e^{-\chk{U}} D^\alpha\left(e^{-\bar{P}}e^{-P}
\bar{D}_\alpha(e^Pe^{\bar{P}})\right) e^{\chk{U}}~,
\end{align}
or in the vector representation
\begin{align}
W^0 = \frac12 e^{-P} \bar{D}^\alpha \left( D_\alpha (e^{P}e^{\bar{P}}) e^{-\bar{P}}e^{-{P}}\right) e^P = \frac12
e^{\bar{P}} D^\alpha\left(e^{-\bar{P}}e^{-P}
\bar{D}_\alpha(e^Pe^{\bar{P}})\right) e^{-\bar{P}}~.
\end{align}

\subsubsection{The action}
The Yang-Mills action is constructed using the projective superfield $\boldsymbol{\mathcal W}$. Since the field is projective the action is integrated with the projective measure $D^2\bar{D}^2$
\begin{align}
I_{N3YM} &= \frac14 \int d^3x D^2\bar{D}^2 \oint \frac{d\zeta}{\zeta} \tr(\boldsymbol{\mathcal W}^2(\zeta)) = \int d^3x D^2\bar{D}^2 \tr(({\mathcal W}^0)^2-\frac12 {\mathcal W}{\mathcal W}^\dg)
\nonumber \\
&= \int d^3x D^2\bar{D}^2 \tr((W^0)^2-\frac12 W{W}^\dg)~.
\end{align}
Using the Bianchi identities (\ref{bianchi}) this can equivalently be written
\begin{align}\label{eq:n3YM}
I_{N3YM} = \int d^3x D^2(D^{0})^{2}\tr W^2 = \int d^3x \bar{D}^2 (D^{0})^{2} \tr({W}^\dg)^2~.
\end{align}
Our next question is what theory the FI term describes. In ${\mathcal N}=3$ in three dimensions, the full measure $d^3xd^6\theta$ is dimensionless. Since the generalized FI term is also dimensionless, there is no need to introduce a dimensionfull coupling constant which indicates that the theory defined in this way could be Chern-Simons theory. To investigate this we now go to components
\begin{align}
S = \int d^3x d^6\theta \sum_{n=1}^\infty \frac{(-1)^{n}}{n}
\oint d\zeta_1\ldots d\zeta_n \tr\left[
\frac{X_1\ldots X_n}{\zeta_{21}\ldots\zeta_{n,n-1}\zeta_{1n}}
\right]~.
\end{align}
Writing the measure $d^6\theta = D^2\bar{D}^2 (D^0)^2 \propto D^2\bar{D}^2\nabla_1^2$ we act with the $\nabla^2$ again using that $X$ is projective
\begin{align}
S = \int d^3x D^2\bar{D}^2 \sum_{n=1}^\infty \frac{(-1)^{n}}{n}
\oint d\zeta_1\ldots d\zeta_n \tr\left[
\frac{X_1\nabla_1^2(X_2\ldots X_n)}{\zeta_{21}\ldots\zeta_{n,n-1}\zeta_{1n}}
\right]~.
\end{align}
We expand the $\zeta$ factors in the same way as for the field strength with the result
\begin{align}
S = \int d^3x D^2\bar{D}^2 \tr\left[
A\bar{D}^2 A + \bar{A}D^2\bar{A} + f_{D,\bar{D}}
\right]~,
\end{align}
where $f_{D,\bar{D}}$ denotes the mixed term with one $D$ and one $\bar{D}$ which will require more work to analyze. Note that the first two terms can be written as superpotentials
\begin{align}
\int d^3 x D^2 \tr (W^2) + \int d^3 x \bar{D}^2 \tr(({W}^\dg)^2)~.
\end{align}
As stated, the mixed term is more complicated. It can be written as
\begin{align}
\int d^3 x D^2\bar{D}^2 \sum_{n=1}^{\infty}\sum_{0\leq k+l\leq n} \frac{(-1)^{n+1}}{n}\tr\left[ R^{(k)} \bar{D}^\alpha ((D_\alpha \Gamma^{(l)}) L^{(n-k-l)})
\right]~,
\end{align}
where
\begin{align}
R^{(k)} = \oint d\zeta_1\ldots d\zeta_k \left(\frac{1}{\zeta_1}+\ldots + \frac{1}{\zeta_k}\right)\frac{X_1\ldots X_k}{\zeta_{21}\ldots\zeta_{k,k-1}}~.
\end{align}
We have not been able to perform the sum but to at least partly analyze the meaning of this term we put all components of $V = \sum_{n=-\infty}^\infty v_n\zeta^n$ to zero except for $v_0$ which is the ${\mathcal N} = 2$ gauge field. Then $X\rightarrow e^{v_0} - 1$ becomes $\zeta$ independent and thus commute with each other. Also, many of the $\zeta$ integrals now vanish, the only nonzero integrals are
\begin{align}\nn
L^{(k)} &\rightarrow X^k\\
\Gamma^{(1)} &\rightarrow X
\end{align}
Using this we can write the mixed term as
\begin{align}\nn
S_{D\bar{D}}(v_0) &= \int d^3x D^2\bar{D}^2 \sum_{n=1}^{\infty}\frac{(-1)^n}{n} \sum_{k=1}^{n} 
\tr(X^{k-1}\bar{D}^\alpha((D_\alpha X)X^{n-k})\\  &=
\int d^3x D^2\bar{D}^2 \sum_{n=1}^{\infty}\frac{(-1)^{n+1}}{n} \sum_{k=1}^{n} 
\tr((\bar{D}^\alpha X^{k-1})(D_\alpha X)X^{n-k})~.
\end{align}
To be able to perform the sum it is useful to write the expression in operator formalism where
\begin{align}
DX \rightarrow [D,X]~.
\end{align}
Then the mixed term becomes
\begin{align}
S_{D\bar{D}}(v_0) &= \int d^3x D^2\bar{D}^2 \sum_{n=1}^{\infty}\frac{(-1)^{n+1}}{n} 
\tr([D^\alpha, X] \sum_{k=1}^{n}  X^{n-k} [\bar{D}_\alpha, X^{k-1}])~,
\end{align}
and writing out the commutators explicitly the sum that we would like to do is
\begin{align}\label{eq:sum}
\sum_{n=1}^{\infty} \frac{(-1)^{n+1}}{n}\left(
\bar{D}_\alpha X^{n-1} + X \bar{D}_\alpha X^{n-2} + \ldots + X^{n-1}\bar{D}_\alpha
- n X^{n-1}\bar{D}_\alpha
\right)~.
\end{align}
In the last term the factor $n$ cancels the $\frac1n$ and gives a geometric series. The first $n$ terms also gives a sumable expression in the following way. If we would replace the $\bar{D}$ with a $\delta X$ the result of the sum would be $\delta \ln(1+X) = \delta v_0$. The idea is then to rewrite $\delta v_0$ in therms of $\delta X = \delta e^{v_0}$ where finally we can change back $\delta X \rightarrow \bar{D}$ for the final result. We do this using the operator ${\mathcal L}$ defined as ${\mathcal L} X = [v_0,X]$ as explained in \cite{Gates:1983nr}
\begin{align}
\delta v_0 = \frac{\mathcal L}{1-e^{-\mathcal L}} e^{-v_0}\delta e^{v_0} =
\frac{\mathcal L}{1-e^{-\mathcal L}} e^{-v_0}\delta X ~.
\end{align}
Following our argument we now conclude that the first $n$ terms sum to $\frac{\mathcal L}{1-e^{-\mathcal L}} e^{-v_0} \bar{D}_\alpha$ and that the full result can be written as
\begin{align}
S_{D\bar{D}}(v_0) = \int d^3x D^2\bar{D}^2 \tr\left(
[D^\alpha,X]\left\{\frac{\mathcal L}{1-e^{-\mathcal L}} e^{-v_0}
- \frac{1}{1+X}\right\} \bar{D}_\alpha 
\right)~,
\end{align}
where the first term in the curly brackets comes from summing the the first $n$ terms of (\ref{eq:sum}) and the second term is the result of the geometric sum from the last term.
Since $1+X = e^{v_0}$ this can be written as
\begin{align}\nn
S_{D\bar{D}}(v_0) &= \int d^3x D^2\bar{D}^2 \tr\left(
[D^\alpha, e^{v_0}]e^{-v_0}\left\{\frac{\mathcal L}{1-e^{-\mathcal L}}
- 1\right\} \bar{D}_\alpha 
\right) \\ &= 
\int d^3x D^2\bar{D}^2 \tr\left(
[D^\alpha,e^{v_0}]e^{-v_0}\left\{\frac{{\mathcal L} - 1 + e^{-\mathcal L}}{1-e^{-\mathcal L}}
\right\} \bar{D}_\alpha 
\right)~.
\end{align}
Similarly we may rewrite $[D^\alpha,e^{v_0}] e^{-v_0} = \frac{e^{\mathcal L} - 1}{\mathcal L} [D^\alpha,v_0]$. Finally, using the cyclicity of the trace implying $\tr(({\mathcal L}B)C)= \tr([v_0,B]C) = -\tr(B[v_0,C])= \tr(B(-{\mathcal L}C))$, where $B$ and $C$ are arbitrary expressions, we write
\begin{align}\nn
S_{D\bar{D}}(v_0) &=
\int d^3x D^2\bar{D}^2 \tr\left(
[D^\alpha,v_0] \frac{1-e^{-\mathcal L}}{\mathcal L}
\frac{{\mathcal L} - 1 + e^{-\mathcal L}}{1-e^{-\mathcal L}}
 \bar{D}_\alpha 
\right) \\  &=
\int d^3x D^2\bar{D}^2 \tr\left(
[D^\alpha,v_0] 
\frac{{\mathcal L} - 1 + e^{-\mathcal L}}{\mathcal L}
 \bar{D}_\alpha 
\right) \\ &=
-\int d^3x D^2\bar{D}^2 \tr\left(
[D^\alpha,v_0] 
\frac{{\mathcal L} - 1 + e^{-\mathcal L}}{{\mathcal L}^2}
 [\bar{D}_\alpha , v_0]\nn
\right)
\end{align}
which, if we go back to standard notation $[\bar{D}_\alpha,v_0] \rightarrow \bar{D}_\alpha v_0$ and expand the function of ${\mathcal L}$ in a power series can be written
\begin{align} S_{D\bar{D}}(v_0)  =
-\int d^3x D^2\bar{D}^2 \tr\left(
\frac12 D^\alpha v_0 \bar{D}_\alpha v_0
- \frac16 D^\alpha v_0 [v_0 , \bar{D}_\alpha v_0]
+ \frac{1}{24} D^\alpha v_0 [v_0, [v_0 , \bar{D}_\alpha v_0]]
+\ldots
\right)
\end{align}
which is indeed a known form for the action of ${\mathcal N} = 2$ Chern-Simons theory \cite{Ivanov:1991fn,Gates:1991qn,Nishino:1991sr}. We thus conclude that the action given by the generalized FI term, which in other theories gives the action for Yang-Mills theory, in the ${\mathcal N}=3$ theory surprisingly gives the action for Chern-Simons theory.

\subsection{${\cal N}=4$ in three dimensions}
This topic has appeared in the literature before. The fact that there are two inequivalent vector multiplets was observed in \cite{Brooks:1994nn} and was later treated in an off-shell fashion in Harmonic superspace in \cite{Zupnik:1999iy}. It was subsequently discussed in Projective superspace in \cite{Kuzenko:2011xg,Kuzenko:2014mva} in an $AdS_3$ background. 

\subsubsection{The algebra}
The ${\mathcal N}=4$ algebra has an $SO(4) \equiv SU(2)_L\times SU(2)_R$ R-symmetry. The supercovariant derivatives transforms in the fundamental representation of each of the R-symmetry groups. The most general algebra involving also a Yang-Mills field is
\begin{align}
    \{{\mathcal D}^{ia}_\alpha , {\mathcal D}^{kb}_\beta\} = \epsilon^{ik}\epsilon^{ab}i\nabla_{\alpha\beta} +
    iC_{\alpha\beta}\left(\epsilon^{ik}W_R^{ab}+\epsilon^{ab}W_L^{ik}\right)~,
\end{align}
where $W_L^{ik}$ and $W_R^{ab}$ are symmetric $SU(2)_L$ and $SU(2)_R$ triplets. The Bianchi identities are
\begin{align}\nn
    {\mathcal D}_\alpha^{i(a}W_R^{bc)} &= 0\\
    {\mathcal D}_\alpha^{(i|a|}W_L^{kl)} & = 0~.
\end{align}

The ${\mathcal N}=4$ theory is the dimensional reduction of the ${\mathcal N}=2$ theory in four dimensions. With the identification
\begin{align}\nn
    {\mathcal D}_\alpha = {\mathcal D}^{11}_\alpha &&
    {\mathcal Q}_\alpha = i{\mathcal D}^{21}_\alpha\\
    \bar{\mathcal Q}_\alpha = i{\mathcal D}^{12}_\alpha &&
    \bar{\mathcal D}_\alpha = {\mathcal D}^{22}_\alpha~,
\end{align}
and
\begin{align}\nn
  {W}_L^\dg = W^{11}_L \;\;\; W_L = W^{22}_L \;\;\; W_L^0 = iW^{12}_L\\
  {W}_R^\dg = W^{11}_R \;\;\; W_R = W^{22}_R \;\;\; W_R^0 = iW^{12}_R~,
\end{align}
the algebra becomes
\begin{align}
\nn
    \left\{{\mathcal D}_\alpha,\bar{\mathcal D}_\beta\right\} =
    i\nabla_{\alpha\beta} &+ C_{\alpha\beta}(W_R^{0}+W_L^{0})\\
    \left\{{\mathcal Q}_\alpha,\bar{\mathcal Q}_\beta\right\} =
    i\nabla_{\alpha\beta} &+ C_{\alpha\beta}(W_R^{0}-W_L^{0})\\ \nn
    \left\{{\mathcal D}_\alpha,{\mathcal Q}_\beta\right\} = -C_{\alpha\beta}{W}^\dg_R \; ; & \;\;
    \left\{\bar{\mathcal D}_\alpha,\bar{\mathcal Q}_\beta\right\} = C_{\alpha\beta}W_R\\
    \left\{{\mathcal D}_\alpha,\bar{\mathcal Q}_\beta\right\} = -C_{\alpha\beta}{W}^\dg_L \; ; & \;\;
    \left\{\bar{\mathcal D}_\alpha,{\mathcal Q}_\beta\right\} = C_{\alpha\beta}W_L~,\nn
\end{align}
and the Bianchi identities for the ${\mathbb W}_R$ multiplet become
\begin{align}\label{eq:BianchiL}
\nn
{\mathcal D}_\alpha {W}_R^\dg = {\mathcal Q}_\alpha {W}_R^\dg &= 0\\
\nn
\bar{\mathcal D}_\alpha {W}_R = \bar{\mathcal Q}_\alpha {W}_R &= 0\\
{\mathcal D}_\alpha {W}_R - 2\bar{\mathcal Q}_\alpha W^0_R &= 0\\
\nn
\bar{\mathcal Q}_\alpha {W}^\dg_R + 2{\mathcal D}_\alpha W^0_R &= 0\\
\nn
\bar{\mathcal D}_\alpha {W}^\dg_R - 2{\mathcal Q}_\alpha W^0_R &= 0\\
\nn
{\mathcal Q}_\alpha {W}_R + 2\bar{\mathcal D}_\alpha W^0_R &= 0~,
\end{align}
and for the ${\mathbb W}_L$ multiplet we get
\begin{align}\label{eq:BianchiR}
\nn
{\mathcal D}_\alpha {W}^\dg_L = \bar{\mathcal Q}_\alpha {W}^\dg_L &= 0\\
\nn
\bar{\mathcal D}_\alpha {W}_L = {\mathcal Q}_\alpha {W}_L &= 0\\
{\mathcal D}_\alpha {W}_L - 2{\mathcal Q}_\alpha W^0_L &= 0\\
\nn
{\mathcal Q}_\alpha {W}^\dg_L + 2{\mathcal D}_\alpha W^0_L &= 0\\
\nn
\bar{\mathcal D}_\alpha {W}^\dg_L - 2\bar{\mathcal Q}_\alpha W^0_L &= 0\\
\nn
\bar{\mathcal Q}_\alpha {W}_L + 2\bar{\mathcal D}_\alpha W^0_L &= 0~.
\end{align}
The fact that the $R$-symmetry group now consists of two $SU(2)$ groups $SU(2)_L\times SU(2)_R$ leads us to introduce two auxilliary $\cpone$ manifolds $\cpone_L \times \cpone_R$ with coordintes $\zeta_L$ and $\zeta_R$.
It is then natural to define the projective supercovariant derivatives
\begin{align}
\nn
    \nabla_\alpha &= ({\mathcal D}_\alpha + \zeta_L {\mathcal Q}_\alpha) + \zeta_R(\bar{\mathcal Q}_\alpha - \zeta_L\bar{\mathcal D}_\alpha)\\
    \label{eq:N4nabla}
    \triL_\alpha &= ({\mathcal D}_\alpha + \zeta_L {\mathcal Q}_\alpha) + \zeta_R(-\bar{\mathcal Q}_\alpha + \zeta_L\bar{\mathcal D}_\alpha)\\
    \nn
    \Delta_\alpha &= (\bar{\mathcal D}_\alpha + \frac{1}{\zeta_L}\bar{\mathcal Q}_\alpha) + \frac{1}{\zeta_R}({\mathcal Q}_\alpha - \frac{1}{\zeta_L}{\mathcal D}_\alpha)\\ \nn
    \triR_\alpha &= (\bar{\mathcal D}_\alpha + \frac{1}{\zeta_L}\bar{\mathcal Q}_\alpha) + \frac{1}{\zeta_R}(-{\mathcal Q}_\alpha + \frac{1}{\zeta_L}{\mathcal D}_\alpha)~.
\end{align}
Even when taken at the same point at both $\cpone_L$ and $\cpone_R$ the gauged supercovariant derivatives do not anticommute. Instead the anticommutation relations are
\begin{align}
\nn
\{\nabla_\alpha, \triL_\beta\} &= 2 C_{\alpha\beta} \zeta_L\zeta_R 
\left(\frac{1}{\zeta_L}{W}^\dg_L + 2W^0_L - \zeta_L W_L\right) =
2 C_{\alpha\beta} \zeta_L\zeta_R {\mathbb W}_L(\zeta_L)\\
\{\nabla_\alpha, \triR_\beta\} &= 2C_{\alpha\beta} \left(
\frac{1}{\zeta_R}{W}^\dg_R +2 W^0_R -\zeta_R W_R\right) = 
2C_{\alpha\beta} {\mathbb W}_R(\zeta_R)\\
\nn
\{\Delta_\alpha, \triL_\beta\} &= 2C_{\alpha\beta} \left(
\frac{1}{\zeta_R}{W}^\dg_R - 2 W^0_R -\zeta_R W_R\right)\\
\nn
\{\Delta_\alpha, \triR_\beta\} &= \frac{2 C_{\alpha\beta}}{\zeta_L\zeta_R} 
\left(\frac{1}{\zeta_L}{W}^\dg_L - 2W^0_L - \zeta_L W_L\right)~,
\end{align}
where we have defined the ${\mathcal O}(2)$ field strengths
\begin{align}\label{eq:PFSR}
{\mathbb W}_R(\zeta_R) &= \frac1{\zeta_R} {W}^\dg_R + 2 W^0_R -\zeta_R W_R\\ \nn
{\mathbb W}_L(\zeta_L) &= \frac1{\zeta_L} {W}^\dg_L + 2 W^0_L -\zeta_L W_L~.
\end{align}

Furthermore, using these definitions, the Bianchi identities can be compactly written as
\begin{align}
\nabla_\alpha {\mathbb W}_L &= \triL_{\alpha} {\mathbb W}_L = 0\\ \nn
\nabla_\alpha {\mathbb W}_R &= \triR_\alpha {\mathbb W}_R = 0~.
\end{align}

Notice that the projective derivatives $\nabla_\alpha, \triL_\alpha$ form an anticommuting set only if ${\mathbb W}_L = 0$ and that $\nabla_\alpha, \triR_\alpha$ anticommute only if ${\mathbb W}_R = 0$. Thus, if and only if ${\mathbb W}_L = 0$ we may consistently define left projective superfields $T_L$ as fields satisfying the constraint
\begin{align}
\nabla_\alpha T_L = \triL_\alpha T_L = 0~,
\end{align}
and similarly, if and only if ${\mathbb W}_R = 0$ we may consistently define right projective superfields $T_R$ as fields satisfying
\begin{align}
\nabla_\alpha T_R = \triR_\alpha T_R = 0~.
\end{align}
This theory was studied before in Projective superspace \cite{Arai:2012dn,Arai:2013eta} but in $\mathcal{N}=4$ unitary gauge.

\subsubsection{The field strength}
If we assume the ${\mathbb W}_L = 0$ so that we may consistently define left projective matter superfields $\Upsilon_L(\zeta_L)$, independent of $\zeta_R$ and with polar dependence on $\zeta_L$
\begin{align}
\nabla_\alpha \Upsilon_L = \triL_\alpha \Upsilon_L = 0~.
\end{align}
Such fields transform under gauge transformations as
\begin{align}
\Upsilon_L \rightarrow e^{i\Lambda_L}\Upsilon_L~,
\end{align}
where also $\Lambda_L$ is left projective. In this case the gauge field $e^{V_L}$ is introduced as a tropical left projective superfield transforming under gauge transformations as
\begin{align}
e^{V_L} \rightarrow e^{i\bar{\Lambda}_L}e^{V_L}e^{-i\Lambda_L}~.
\end{align}
We can repeat what we did in the previous sections by observing that if we take the anticommutator of two $\nabla$ derivatives at different $\zeta_L$ positions we have
\begin{align}
\{\nabla_\alpha(\zeta_{1L},\zeta_R), \nabla_\beta(\zeta_{2L},\zeta_R)\} =
(\zeta_{1L}-\zeta_{2L})C_{\alpha\beta} \zeta_R {\mathbb W}_R(\zeta_R)~,
\end{align}
which, introducing the derivative operator $\partial_L \equiv \frac{\partial}{\partial \zeta_L}$, can also be written
\begin{align}
\{[\partial_L,\nabla_\alpha(\zeta_L,\zeta_R)],\nabla_\beta(\zeta_L,\zeta_R)\} =
C_{\alpha\beta} \zeta_R {\mathbb W}_R(\zeta_R)~.
\end{align}
Splitting the gauge field into polar parts
\begin{align}
e^{V_L} = e^{\bar{U}_L}e^{U_L}~,
\end{align}
we define the field strength in arctic representation
\begin{align}
\{[e^{-U_L}\partial_L e^{U_L},\nabla_\alpha(\zeta_L,\zeta_R)], \nabla_\beta(\zeta_L,\zeta_R)\} =
C_{\alpha\beta} e^{-U_L} \zeta_R {\mathbb W}_R(\zeta_R) e^{U_L} = C_{\alpha\beta}
\zeta_R \boldsymbol{\mathcal W}_R(\zeta_L,\zeta_R)~.
\end{align}
We see that in the arctic representation, the field strength is an ${\mathcal O}(2)$ multiplet in $
\zeta_R$ but has an arctic dependence on $\zeta_L$.
By defining the $\zeta_L$ covariant derivative
\begin{align}
{\cal D}_L = e^{-U_L}\partial_L e^{U_L} = \partial_L + {\mathcal A}_L~,
\end{align}
we can write the field strength in the arctic representation as
\begin{align}
\nn
\boldsymbol{\mathcal W}_R &= -\frac{1}{\zeta_R}\nabla^2 {\mathcal A}_L =
\frac{1}{\zeta_R}{\mathcal W}_R^\dg + 2{\mathcal W}_R^0 - \zeta_R{\mathcal W}_R = \\
&-\left(\frac{1}{\zeta_R} (D+\zeta_L Q)^2 +(D+\zeta_L Q)^\alpha (\bar{Q} - \zeta_L \bar{D})_\alpha + \zeta_R (\bar{Q}-\zeta_L \bar{D})^2 \right) {\mathcal A}_L~,
\end{align}
giving the component field strengths
\begin{align}
\nn
{\mathcal W}^\dg_R &= -(D+\zeta_L Q)^2 {\mathcal A}_L\\
2{\mathcal W}^0_R &= - (D+\zeta_L Q)^\alpha (\bar{Q} - \zeta_L \bar{D})_\alpha {\mathcal A}_L\\
\nn
{\mathcal W}_R &= (\bar{Q}-\zeta_L \bar{D})^2 {\mathcal A}_L~.
\end{align}
The new techniques introduced in this paper allow us to find the components of these expressions. We find that
\begin{align}\label{eq:n4comp}
\nn
{\mathcal W}^\dg_R(\zeta_L) &= e^{-\chk{U}_L} (D^2 \bar{A}) e^{\chk{U}_L}
\\
2{\mathcal W}^0_R(\zeta_L) &= -e^{-\hat{U}_L}\bar{D}^\alpha(D_\alpha(e^Pe^{\bar{P}})e^{-\bar{P}}e^{-P})e^{\hat{U}_L} =
- e^{-\chk{U}_L} D^\alpha (e^{-\bar{P}}e^{-P} \bar{D}_\alpha (e^{P}e^{\bar{P}})) e^{\chk{U}_L}
\\ \nn
{\mathcal W}_R(\zeta_L) &=  e^{-\hat{U}_L} (\bar{D}^2 A )e^{\hat{U}_L}~,
\end{align}
where the fields depend on $\zeta_L$. Notice that the expressions for ${\mathcal W}_R,{\mathcal W}^\dg_R$ are exactly the same as for the field strength in four dimensions.

An entirely analogous argument starting with ${\mathbb W}_R = 0$ and right projective matter superfields gives us the expression for the field strength in the right projective sector as
\begin{align}
\nn
\boldsymbol{\mathcal W}_L &= -\frac{1}{\zeta_L}\nabla^2 {\mathcal A}_R
= \frac{1}{\zeta_L}{\mathcal W}^\dg_L+2{\mathcal W}^0_L - \zeta_L{\mathcal W}_L\\
{\mathcal A}_R &= e^{-U_R}(\partial_R e^{U_R})~,
\end{align}
leading to the expressions for the components of the field strengths
\begin{align}
\nn
{\mathcal W}^\dg_L &= -(D+\zeta_R\bar{Q})^2 {\mathcal A}_R
\\
2{\mathcal W}^{0}_L &= - (D +\zeta_R\bar{Q})^\alpha (Q-\zeta_R\bar{D})_\alpha {\mathcal A}_R
\\ \nn
{\mathcal W}_L &= (Q - \zeta_R \bar{D} )^2 {\mathcal A}_R~.
\end{align}
The expressions for the field strengths when expressed using only the $D$ and $\bar{D}$ operators are identical to (\ref{eq:n4comp}) but with all the expressions computed from $e^{U_R}$ instead of $e^{U_L}$.

\subsubsection{The action}
Using the projective field strength (\ref{eq:PFSR}) we can write a gauge invariant action using the projective measure
\begin{align}
S = \int d^3 x D^2\bar{D}^2 \oint \frac{d\zeta_R}{\zeta_R} \tr ({\mathbb W}_R^2) =
\int d^3 x D^2\bar{D}^2 \tr (4(W_R^0)^2-2W_R {W}^\dg_R)~.
\end{align}
If we let the $\bar{D}^2$ derivatives act and use the Bianchi identities (\ref{eq:BianchiL}) the action can be rewritten in chiral form as
\begin{align}
S = \int d^3 x D^2Q^2 \tr(W_R^2)~,
\end{align}
where $W_R$ is the component of ${\mathbb W}_R$ that multiplies $\zeta$.
To show that the generalized Fayet-Iliopoulos term is equivalent to this action we follow exactly the same steps as in the four dimensional case. We write the full superspace measure as $D^2Q^2\bar{D}^2\bar{Q}^2 = D^2Q^2\bar{D}^2(\bar{Q} - \zeta \bar{D})^2$ and use that since a left-projective superfield is annihilated by both $\nabla_\alpha$ and $\triL_\alpha$ it is also annihilated by $\frac{1}{2\zeta_R} (\nabla_\alpha -\triL_\alpha) = (\bar{Q}-\zeta\bar{D})_\alpha$. The formulas are precisely the formulas from the four dimensional example (\ref{eq:n2action}) with the same result (\ref{eq:n2YM}) showing that the generalized Fayet-Iliopoulos term again gives the action of Yang-Mills theory.
An analogous story holds in the right-projective case.

It is possible to reduce expressions in ${\mathcal N}=4$ superspace to ${\mathcal N}=3$ superspace by rewriting the complex supercovariant derivative $Q_\alpha$ in terms of real and imaginary part as
\begin{align}
Q_\alpha = D^{0}_\alpha +i Q^{0}_\alpha~,
\end{align}
and then use that on left projective superfields, after setting $\zeta_L=\zeta_R = \zeta$
\begin{align}
Q^{0}_\alpha = \frac{1}{\zeta} D_\alpha + \zeta \bar{D}_\alpha~.
\end{align}
Starting from the Yang-Mills action, if we expand the $\zeta$ factors according to the rules developed in this paper, we end up with the ${\mathcal N}=3$ action for Yang-Mills theory (\ref{eq:n3YM}).

\section{Conclusions}
We introduced a new method for manipulating gauge fields in Projective superspace. The main idea was to expand all expressions in powers of $X = e^V-1$ (and thus indirectly in powers of the unconstrained gauge prepotential $V$) and then to use the epsilon prescription when projecting fields to positive or negative powers of $\zeta$ leading to an asymmetric splitting which, however, is much more convenient to calculate with. This new approach led to elegant formulas for gauge potentials and field strengths. In particular, it becomes straightforward to rewrite the expressions in components, i.e. in a superspace with less manifest supersymmetry (${\mathcal N} =1$ in four dimensions, ${\mathcal N}=2$ in three dimensions) which also made it possible to analyze various superspace Lagrangians in detail. We illustrated our results in several explicit examples. 

Using the techniques introduced in this paper we get formulations of a wide range of theories in terms of the unconstrained gauge prepotential $V$. This makes it possible to calculate perturbative quantum corrections to these theories keeping the supersymmetry manifest.

We have not investigated the dimensional reduction of our theories to two dimensions. In two dimensions there are even more possibilities for Projective techniques since a theory with $(p,q)$ supersymmetry has a $SO(p)\times SO(q)$
R-symmetry group. For a theory with $(4,4)$ supersymmetry that means that four auxilliary $\cpone$ manifolds could be introduced \cite{Ivanov:1991fn}. It would be interesting to see how the different ${\mathcal N} = 3$ and ${\mathcal N} = 4$ theories fit into this picture.

Another area where the methods from this paper should be useful is in studying mirror symmetry of three dimensional ${\mathcal N}=4$ supersymmetric theories \cite{Intriligator:1996ex}. The mirror symmetry exchanges a left projective theory with a (twisted) right projective theory through a duality procedure that can be explicitly performed in the path integral. The coupling between the original theory and the twisted theory is through a BF type term which we know how to write in our formalism. Since both the Coulomb and the Higgs branches of the mirror symmetric theories are Hyperkähler manifolds, this might lead to new constructions of and relations between Hyperkähler manifolds.

Similarly our methods are well suited to investigate ABJM theory \cite{Aharony:2008ug} in superspace since there one describes the ${\mathcal N}=6$ supersymmetric theory in terms of two ${\mathcal N}=3$ Chern-Simons theories together with a pair of bifundamental hypermultiplets. This was first done in ${\mathcal N}=3$ superspace in \cite{Buchbinder:2008vi}. The possibility to perform perturbative calculations with manifest ${\mathcal N}=3$ supersymmetry and to extensively analyze the result should lead to new insights.

Another situation where our newly developed procedures may come in handy is for Yang-Mills theories, or in particular Chern-Simons theory, in five dimensions \cite{Zupnik:1999iy,Kuzenko:2005sz,Kuzenko:2006ek}. In \cite{Kuzenko:2006ek} an action for five dimensional Chern-Simons was given in the abelian case whereas for the non-abelian only the variation of the action was given. We believe that using the techniques introduced in this paper and suitably adapted to five dimensions will allow us to integrate this variation and find a covariant expression for the full nonabelian Chern-Simons action.

\begin{acknowledgments}
The authors thank Martin Roček and Linus Wulff for useful discussions.
The work of Rikard von Unge is supported by the Czech science foundation GA\v{C}R through the grant “Integrable Deformations” (GA20-04800S) from 2020-2022 and currently from the grant "Dualities and higher derivatives" (GA23-06498S).
Ulf Lindstr\"om is supported by the Leverhulme trust through a Leverhulme Visiting professorship.
\end{acknowledgments}

\appendix

\section{The epsilon prescription\label{app:eps}}
\subsection{Definition}
In the paper we employ a regularization method of the poles appearing in some of the contour integrations. It was originally introduced in \cite{Jain:2009aj,Jain:2010gm} and called the $\epsilon$-prescription. Following \cite{Jain:2009aj,Jain:2010gm} we redefine
\begin{align}
\frac{1}{\zeta_1-\zeta_2} \rightarrow \frac{1}{\zeta_{12}}~,
\end{align}
where
\begin{align}\label{eq:epdef}
\frac{1}{\zeta_{10}} &\equiv 
\frac{1}{\zeta_1} \sum_{n=0}^{\infty} \left(\frac{\zeta_0}{\zeta_1}\right)^n 
\left(= \frac{1}{\zeta_1-\zeta_0} \;\;\;{\rm if}\;\;\; \left|\frac{\zeta_0}{\zeta_1}\right|<1\right) \\ \nonumber
\frac{1}{\zeta_{01}} &\equiv
\frac{1}{\zeta_0} \sum_{n=0}^{\infty} \left(\frac{\zeta_1}{\zeta_0}\right)^n \left(=
\frac{1}{\zeta_0-\zeta_1} \;\;\;{\rm if} \;\;\; \left|\frac{\zeta_1}{\zeta_0}\right|<1 \right)~.
\end{align}
Alternatively, we may define the factors using a small parameter $\epsilon$ (hence the name) that we will let go to zero at the end of all calculations
\begin{align}
\frac{1}{\zeta_{12}} = \frac{1}{\zeta_1-\zeta_2+\epsilon(\zeta_1+\zeta_2)}~.
\end{align}
In this expression $\zeta_1$ and $\zeta_2$ should be thought of as having the same magnitude $|\zeta_1| = |\zeta_2|$ so that the $\zeta_1$ contour would actually meet $\zeta_2$. By introducing the $\epsilon$, the pole is deformed away from the integration contour and the integral can be performed. The $\epsilon$-regulated denominators give rize to projection operators, more precisely, given a superfield $T(\zeta) = \sum_{n=-\infty}^{\infty} t_n\zeta^n$ the contour integrals are
\begin{align}\nonumber
\oint \frac{d\zeta_1}{2\pi i} \frac{T(\zeta_1)}{\zeta_{10}} &=
\oint \frac{d\zeta_1}{2\pi i}  \frac{1}{\zeta_1} \sum_{n=0}^{\infty} \left(\frac{\zeta_0}{\zeta_1}\right)^n T(\zeta_1)  = \sum_{k=0}^{\infty} t_k \zeta_0^k\\ \label{eq:eppr} 
\oint \frac{d\zeta_1}{2\pi i} \frac{T(\zeta_1)}{\zeta_{01}} &=
\oint \frac{d\zeta_1}{2\pi i}  \frac{1}{\zeta_0} \sum_{n=0}^{\infty} \left(\frac{\zeta_1}{\zeta_0}\right)^n T(\zeta_1) = \sum_{k=-\infty}^{-1} t_k \zeta_0^k~,
\end{align}
Notice that the projection defined in this way does not treat the $\zeta$-independent terms in a symmetrical way. They are always projected together with the positive powers of $\zeta$ and thus $\oint \frac{d\zeta_1}{2\pi i} \frac{T(\zeta_1)}{\zeta_{10}}$ and $\oint \frac{d\zeta_1}{2\pi i} \frac{T(\zeta_1)}{\zeta_{01}}$ are not conjugates of each other. To avoid cluttering we will suppress any factors of $2\pi i$ in the measure, $\frac{d\zeta}{2\pi i}\rightarrow d\zeta$, in the rest of the paper.

\subsection{Cancellation}
In several places we cancel factors of $(\zeta_1-\zeta_2)$ against $\epsilon$-regulated denominators $\frac{1}{\zeta_{12}}$. To prove that this is possible we explicitly calculate the integral
\begin{align}
I = \oint d\zeta_1 \frac{\zeta_1-\zeta_2}{\zeta_{12}} T(\zeta_1)~,
\end{align}
for an arbitrary field $T(\zeta) = \sum_{n=-\infty}^{\infty} t_n \zeta^n$. We start by splitting the integral into two integrals
\begin{align}
I  = \oint d\zeta_1 \frac{\zeta_1}{\zeta_{12}} T(\zeta_1) -
\oint d\zeta_1 \frac{\zeta_2}{\zeta_{12}} T(\zeta_1) =
\sum_{n=-1}^{\infty} t_n \zeta_2^{n+1} - \sum_{n=0}^{\infty} t_n \zeta_2^{n+1} = t_{-1}~,
\end{align}
which is precisely the result for the integral $I$ one would get by canceling the $(\zeta_1-\zeta_2)$ against the $\frac{1}{\zeta_{12}}$ before performing the contour integral. An analogous argument can be made for the factor $\frac{1}{\zeta_{21}}$ in the denominator.

\subsection{Conjugation}
The basic operation of conjugation is always accompanied by a transformation on the auxiliary $\cpone$ manifold. A general field $T(\zeta) = \sum_n t_n \zeta^n$ transforms as
\begin{align}
\bar{T}(\zeta) = \sum t_n^{\dagger} \left(-\frac{1}{\zeta}\right)^n~.
\end{align}
An epsilon-regulated $\zeta$ denominator transforms as
\begin{align}
\frac{1}{\zeta_{21}} = \frac{1}{\zeta_2}\sum_{n=0}^\infty \left(\frac{\zeta_1}{\zeta_2}\right)^n \rightarrow
-\frac{\zeta_1\zeta_2}{\zeta_1}\sum_{n=0}^\infty \left(\frac{\zeta_2}{\zeta_1}\right)^n =
-\frac{\zeta_1\zeta_2}{\zeta_{12}}~,
\end{align}
and the contour integral measure transforms as
\begin{align}
\oint d\zeta \rightarrow -\oint \frac{d\zeta}{\zeta^2}~.
\end{align}
Notice the additional minus sign from a change in orientation of the contour.

\bibliographystyle{unsrtnat}


\end{document}